\documentclass[10pt,letterpaper]{IEEEtran}
\IEEEoverridecommandlockouts
   \usepackage{algorithm}
   \usepackage{algpseudocode}
   \usepackage{amsmath}
   \usepackage{amssymb}

\usepackage{subcaption}
\usepackage{bbm}
\usepackage{adjustbox}
\usepackage{lipsum}
\usepackage{stmaryrd}
\usepackage{bm}
\usepackage{cite}
\usepackage{amsmath,amssymb,amsfonts}
\usepackage[compatibility=false]{caption}
\usepackage{graphicx}
\usepackage{siunitx}
\usepackage{textcomp}
\usepackage{tikz}
\usetikzlibrary{positioning,arrows.meta}
\usetikzlibrary{arrows.meta,positioning,calc}
\usepackage{xcolor}
\usepackage[margin=0.7in]{geometry}
\def\BibTeX{{\rm B\kern-.05em{\sc i\kern-.025em b}\kern-.08em
    T\kern-.1667em\lower.7ex\hbox{E}\kern-.125emX}}
\usepackage{amsthm}
\usepackage{amsmath}
\usepackage{amssymb}
\newtheorem{theorem}{Theorem}

\newtheorem{proposition}[theorem]{Proposition}

\begin{document}
\title{\huge Risk-Aware Handover Management Approach for Non-Terrestrial Networks with the NTN-Connect Simulator
\\
    }

\author{Mohammed Almekhlafi, Antoine Lesage-Landry, and Gunes Karabulut Kurt
\thanks{M. Almekhlafi and G. Karabulut~Kurt are with the Department of Electrical Engineering, Polytechnique Montreal \& Poly-Grames Research Centre, Montreal, QC H3T 1J4, Canada (e-mail: \texttt{\{mohammed.al-mekhlafi, gunes.kurt\}@polymtl.ca}).
}
\thanks{A. Lesage-Landry is with the Department of Electrical Engineering, Polytechnique Montreal, Poly-Grames Research Centre, Mila \& GERAD, Montreal, QC H3T 1J4, Canada (e-mail: \texttt{antoine.lesage-landry@polymtl.ca})
}}

\maketitle
\IEEEpeerreviewmaketitle
\begin{abstract}
 The continuous movement of low Earth orbit (LEO) satellites in non-terrestrial networks (NTNs) leads to frequent handovers, requiring joint optimization of visibility duration, transmit power, and user quality-of-service (QoS). This work address this issue by proposing a risk-aware handover management framework that minimizes transmit power and maximizes admitted users under per-satellite power budgets and user-rate chance constraints, while a preference score term penalizes short duration associations. We decompose the resulting mixed-integer nonlinear program (MINLP) into an inner power allocation subproblem and an outer generalized assignment association subproblem. The inner subproblem reformulates the chance constraint into its deterministic counterpart using the value-at-risk (VaR) closed-form. The obtained transmit power allocation is then used in the association subproblem, which is solved optimally using an off-the-shelf solver. Finally, we employ the polynomial-complexity Hungarian algorithm to solve the relaxed assignment problem and enforce the per-satellite power budget using a knapsack algorithm. Simulations, using the proposed NTN-Connect simulator, on the OneWeb and Starlink constellations show that the proposed low-complexity solution achieves near-optimal performance and better performance compared to the risk-aware maximum stay (RA-MST) and risk-aware maximum received signal received power (RA-MRSRP) baselines. On the Starlink constellation, it increases the inter-handover interval (IHI) from $0.37$ to $1.47$~min, reducing the handover rate by $75\%$. At a maximum-IHI operating point, it reduces the total transmit power by 1.1 dBW (22\%) while increasing the admission rate from 97.4\% to 99.3\%.

\end{abstract}


\section{Introduction}
Non-terrestrial networks (NTNs) have emerged as a key enabler technology for the 6th generation (6G) and beyond wireless communication networks. NTNs are expected to support a wide range of 6G and beyond applications, such as eHealth and public safety, while ensuring service continuity, ubiquitous connectivity, and scalability~\cite{almekhlafi_inequality},\cite{3126825},\cite{survey}. Notably, NTN's goal is the continuous service in scenarios where terrestrial infrastructure is not deployable, e.g., polar regions. NTNs also allow the required scalability to meet the increasing demand by providing seamless global connectivity, additional computational resources, and backhauling alternatives~\cite{almekhlafi_inequality}. This is enabled by the diverse NTN components, including geostationary, highly elliptical, medium, and low Earth orbit regimes. These divers regimes offer a distinct trade-off among coverage footprint, propagation delay, and achievable data rate. Particularly, geostationary Earth orbit (GEO) satellites, at an altitude of $35{,}786$~km, are able to provide wide-area and quasi-static coverage compared to medium Earth orbit (MEO) and low Earth orbit (LEO) satellites, which operate at lower altitudes of $2{,}000$--$35{,}786$~km and below $2{,}000$~km, respectively. In this context, MEO constellations have a larger coverage footprint than LEO constellations, while LEO constellations, such as Starlink~\cite{spacex_starlink} and OneWeb~\cite{oneweb_connectivity}, can offer high data rates with lower delay through mega-constellations of thousands of satellites connected by inter-satellite links (ISLs)~\cite{3126825,survey}.%

To realize this vision, the 3rd Generation Partnership Project (3GPP) has undertaken extensive standardization activities for NTNs. The integration of different NTN components into a unified, multi-layered architecture is a central vision of 6G and beyond networks~\cite{3126825,10409745}. However, unlike terrestrial networks, this integration remains challenging because the topology of NTNs continuously changes over time. Consequently, visibility, routing paths, and resource availability vary dynamically, leading to complex cross-layer resource management challenges. For example, ground users must frequently perform handovers between satellites while maintaining stable associations and efficient radio resource management (RRM). Moreover, the changing ISL and network topologies introduce additional coordination complexity.%

Among these challenges, frequent handovers are among the major challenges in LEO satellite networks. In particular, the high frequency of handovers introduces a significant signalling and coordination overhead while complicating efficient resource management~\cite{AlHraishawi,almekhlafi2026icc}. 
The inherent mobility of satellites continues to make RRM more challenging than in terrestrial networks, making efficient handover management critical for maintaining service continuity, improving resource utilization, and satisfying users' quality-of-service (QoS) requirements \cite{almekhlafi2026icc}. The frequent handovers inherent to LEO satellite networks continuously change the serving satellite and propagation conditions, making timely and accurate channel state information (CSI) difficult to acquire and maintain~\cite{alsenwi2024risk}. Hence, handover optimization schemes that ignore channel uncertainty may fail to provide reliable QoS guarantees. This motivates the development of risk-aware handover optimization models that account for stochastic channel.\vspace{-.1 in}
\subsection{Related Work}
\label{sec:related}
We now review the work related to ours. 
\subsubsection{Handover Management in NTNs}
In response to the handover management challenges in NTNs, recent research has studied this problem from different perspectives. The authors of~\cite{10443403} compare three handover strategies for different constellations, concluding that the optimal handover strategy is constellation-dependent. For example, the carrier-to-interference-and-noise ratio (CINR) approach achieves the best spectral-efficiency, propagation-delay, and handover-rate trade-off for large LEO systems, whereas the closest-satellite solution is more effective for smaller constellations. In~\cite{10751771}, the authors propose a MEO-assisted distributed mobility-management framework with a low-overhead handover solution, in which the MEO layer orchestrates handovers in the LEO network. Handover performance in dense LEO networks is analyzed in~\cite{Zhou2023} using stochastic geometry. The Kuhn--Munkres (KM) algorithm is used in~\cite{10884973} to enhance QoS and manage frequent handovers. The work in~\cite{11068183} develops an energy-aware handover solution for a single-user remote scenario. Reference~\cite{Y_WISEE} proposes a convex optimization framework for jointly minimizing power consumption and handover frequency in LEO constellations, while~\cite{11667348} combines convex power allocation with MARL-based beam association to reduce handovers and improve effective throughput. The authors in~\cite{11131136} propose a deep Q-network (DQN) to coordinate gateway--user interference in LEO constellations, thereby jointly enhancing link quality and handover stability. Reference~\cite{hongtao2025double} presents a clustering-based solution to limit simultaneous handovers in mega-LEO constellations. A multi-objective reinforcement learning (RL) problem is formulated in~\cite{10700698} with the aim of maximizing throughput, limiting handovers, and balancing network load. Reference~\cite{hozayen2022graph_handover} formulates handover management as a shortest-path problem on a time-based graph to jointly optimize handover timing and target selection. A multi-agent RL (MARL)-based conditional handover algorithm is proposed in~\cite{choi2025intelligent} with the aim of reducing unnecessary handovers. Finally, the work in~\cite{11419899} proposes a hierarchical reinforcement learning (HRL) framework for handover management in dense LEO satellite constellations by jointly optimizing handover timing and target satellite selection through a hybrid PPO--DQN architecture. 

\subsubsection{Risk optimization in NTNs}
Recent research has increasingly investigated robust optimization techniques that explicitly account for channel uncertainty, interference, and stochastic network dynamics. Existing studies have primarily focused on robust RRM. For example, a robust energy-efficient beamforming for massive MIMO LEO satellite communication systems under imperfect CSI is studied in~\cite{yliu2022robust}. Alsenwi~\emph{et al.} propose a risk-aware learning framework for beamforming by adopting conditional value-at-risk (CVaR) to model channel uncertainty and combining deep reinforcement learning~\cite{alsenwi2024risk}. The authors of~\cite{huang2025robust} develop a robust Lyapunov optimization framework for routing control in LEO satellite networks under bursty traffic arrivals. The proposed method improves routing robustness against uncertain traffic dynamics. Fang \emph{et al.}~\cite{fang2026rpp} introduce a risk-predictive proactive handover strategy for LEO satellite edge computing to improve the reliability of computational task offloading.

\subsubsection{NTN Simulation Frameworks}
\label{sec:design}
Recently, simulation and emulation platforms for NTNs have evolved rapidly to support the analysis of the increasing complexity of NTNs. Network emulators, such as \texttt{StarryNet}~\cite{StarryNet}, \texttt{OpenSN}~\cite{OpenSN}, \texttt{NetSatBench}~\cite{NetSatBench}, and \texttt{XeOverse}~\cite{xeoverse} 
enable realistic evaluation of network protocols and applications in controlled environments. \texttt{StarryNet} employs Docker-based container virtualization to emulate dynamic satellite networks. \texttt{OpenSN} extends this approach with improved container and virtual-link management and flexible multi-machine deployment for enhanced scalability. The authors of \texttt{NetSatBench} introduce a declarative, epoch-driven orchestration framework for distributed container-based satellite network emulation. \texttt{XeOverse} adopts the \texttt{Mininet} framework~\cite{lantz2010network} with updated ISL information and link capacity estimation to achieve real-time satellite network emulation. These emulators prioritize network-software fidelity over physical-layer simulation by modelling wireless links as network abstractions. 

Additionally, several simulators have been introduced for large constellations while capturing topology dynamics \texttt{Hypatia}~\cite{hypatia}, \texttt{StarPerf}~\cite{starperf}, \texttt{SILLEO-SCNS}~\cite{bskempton}, and \texttt{SkyLinkSimulator}~\cite{skylink}. \texttt{Hypatia} integrates Python-based pre-processing of satellite geometry, topology, and routing state with the ns-3 packet-level simulator~\cite{riley2010ns3}. \texttt{StarPerf} is a Python-based simulator that models the time-varying topology of LEO mega-constellations to characterize network performance and explore constellation design trade-offs. \texttt{SILLEO-SCNS} provides graph visualization and routing analysis for LEO constellations. \texttt{SkyLinkSimulator} is a graph-based simulator for ISL-enabled LEO networks and supports distributed routing and link-management algorithms.
\subsubsection{Research Gap}
 Existing handover schemes optimize expected performance metrics, such as throughput, and handover frequency, without accounting for the tail behaviour of the achievable-rate distribution under channel uncertainty. Moreover, in most work, handover decisions are decoupled from admission control and transmit power allocation, which may lead to associations that are not feasible under satellite resource constraints. In parallel, existing risk-aware RRM frameworks mainly focus on isolated optimization problems, including routing, power control, and task scheduling. The joint optimization of user association, admission control, handover management, and transmit power allocation under stochastic QoS guarantees remains unexplored. This work develops a computationally efficient risk-aware optimization model based on chance-constrained programming to jointly optimize user association, admission control, and transmit power while enabling real-time handover management in LEO networks.\vspace{-.13 in}%
\subsection{Contributions}
\label{sec:contributions}
 In this work, we propose a centralized risk-aware handover management framework for LEO networks. In this framework the user association, transmit power allocation, and admission control are jointly optimized under stochastic QoS constraints arising from channel uncertainty. Additionally, we introduce \texttt{NTN-Connect}, a modular, open-source, and configuration-driven simulation framework for NTNs. This tool enables integration of communication, networking, and optimization algorithms through interchangeable components. The main contributions of this paper are as follows.%
\begin{itemize}
\item We propose a risk-aware handover optimization framework that jointly minimizes the handover frequency and transmit power while maximizing user admission. The problem is subject to satellite's power budget and visibility, beam-capacity constraints, and chance rate QoS requirements. To achieve this objective, we introduce a generalized handover-aware association preference (GHAAP) score that jointly balances link quality, remaining visibility duration, and association persistence. This score enables a robust and a stable user--satellite association decisions.

\item We decompose the resulting MINLP problem into an inner power allocation subproblem and an outer association subproblem. For a fixed association, we first reformulate the chance constraint into its deterministic equivalent using the VaR and derive a closed-form expression for per-link minimum transmit power. We approximate the composite lognormal-Rician distribution by a lognormal distribution and obtain a computationally tractable reformulation. The association problem is then formulated as an integer linear program (ILP). 
\item We develop a greedy algorithm to further reduce the computational complexity. It first solves a relaxed linear assignment problem and then restores the aggregate satellite power-budget constraint through a per-satellite $0$--$1$ knapsack algorithm. Additionally, we develop two computationally efficient risk-aware handover schemes, namely, risk-aware maximum stay (RA-MST) and risk-aware maximum received signal received power (RA-MRSRP), by extending conventional MST and MRSRP with risk-aware power allocation.
 \item We evaluate the proposed framework through extensive simulations on the OneWeb and Starlink constellations using the developed \texttt{NTN-Connect} simulator- a research-grade, modular simulator for NTNs. The results demonstrate that the proposed framework achieves a balanced trade-off between user admission, transmit power consumption, and handover stability compared with conventional handover schemes for both constellations.
\end{itemize}

\subsection{Paper Organization}
The remainder of this work is organized as follows. Section~II presents the system model and the problem formulation. Section~III discusses the solution approach. Section~IV introduces the \texttt{NTN-Connect} simulator and presents the simulation results. Finally, Section~V concludes this paper.

\section{System Model and Problem Formulation}
In this section, we present the system model and the problem formulation of the proposed risk-aware handover management framework for LEO networks.
\subsection{System Model}\label{subsec:systemmode}
\begin{figure}[t!]
\centering
\includegraphics[width=0.8\columnwidth,draft=false]{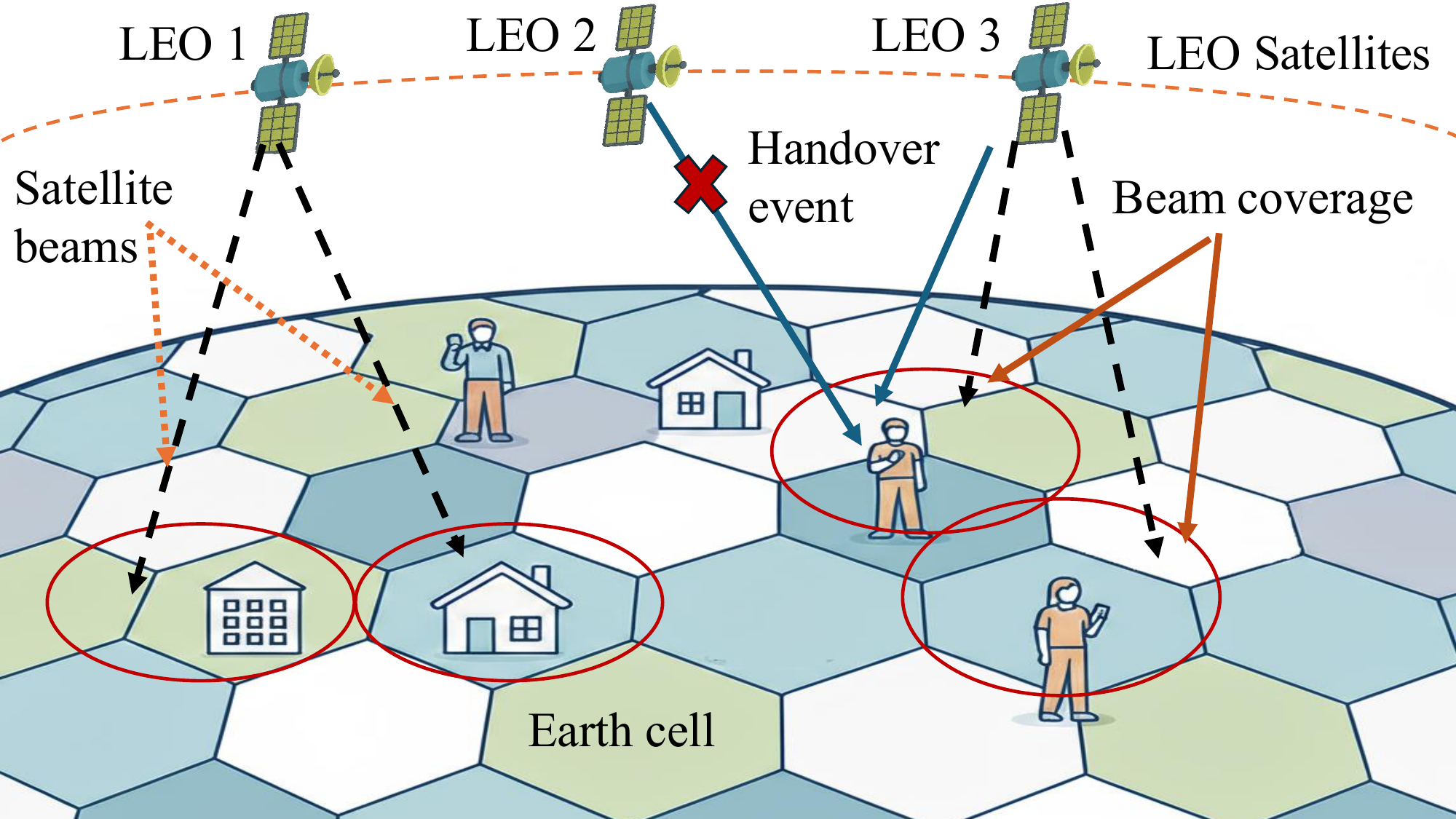}
\caption{System model.}
\label{fig:sys}
\vspace{-.09 in}
\end{figure}

In this work, we consider a mega-constellation LEO satellite network composed of a set of LEO satellites serving a set of ground users as illustrated in Fig.~\ref{fig:sys}. The set of LEO satellites, denoted by $\mathcal{S}=\{1,2,\dots,\rm{S}\}$ are moving in predetermined orbits, where ${\rm S}\in \mathbb{N}$ is the total number of satellites. Additionally, each satellite is equipped with $\rm{B}\in\mathbb{N}$ beams and can simultaneously serve up to $\rm{B}$ ground users. The system bandwidth $W>0$ is divided equally among the beams, so that each beam operates on a bandwidth $W/B$. Accordingly, each beam establishes one communication link with one user, while the total number of active communication links is constrained by the available beams at the satellite. Let $\mathcal{U}=\{1,2,\dots,\rm{U}\}$ be the set of ground users, where ${\rm U}\in \mathbb{N}$ is total number of users. Ground users are distributed across the considered geographical area. We assume that the users are stationary, i.e., the mobility of ground users is negligible relative to the satellite motion.

Because satellites continuously move in the the LEO mega-constellation network, the coverage footprints of multiple satellites overlap. Therefore, a ground user may be visible to multiple satellites at the same time. Let $T\in\mathbb{N}$ denote the total number of time slots in the considered discretized time horizon, with slots indexed by $t=1,2,\dots,T$. Let $\mathcal{S}_{u}(t)\subseteq\mathcal{S}$ denote the set of satellites visible to user $u$ at time slot $t$, where $|\mathcal{S}_{u}(t)|=S_{u}(t)$ is its cardinality. The set of visible satellites at time slot $t$ is defined as $ \mathcal{S}(t)\triangleq \bigcup_{u\in\mathcal{U}}\mathcal{S}_{u}(t) $. The set of visible satellites $\mathcal{S}(t)$ is updated at every time slot based on the instantaneous satellite--user geometry, ensuring that only satellites satisfying the visibility constraint are included. In this regard, we consider that users operate with minimum elevation angles of $\theta^{\min}_{u}$. Accordingly, the visibility indicator between user $u$ and satellite $s$ at time slot $t$ is defined as
\begin{equation}
  V_{u,s}(t)=
\begin{cases}
1, & \theta_{u,s}(t)\geq\theta^{\rm min}_{u},\\
0, & \text{otherwise},
\end{cases}  
\end{equation}
where $\theta_{u,s}(t)$ denotes the elevation angle of satellite $s$ as observed by user $u$ at time $t$. 

The rapid movement of LEO satellites results in continuously varying propagation conditions, making the satellite--user channel inherently stochastic. Accordingly, the satellite--user channel is modelled using the composite propagation model adopted in the standardized 3GPP NTN framework~\cite{3gpp38811}, which jointly captures the large-scale propagation effects and the small-scale fading. Because LEO satellite links are typically dominated by a strong line-of-sight (LoS) component, the small-scale fading coefficient $\tilde{h}_{u,s}(t)$ is modelled as a Rician random variable with elevation-dependent Rician factor $K_{u,s}(t)$. For the satellite--user link, the channel is described by
\begin{equation}\label{eq:channel}
h_{u,s} (t)=\sqrt{G_{\rm rx}G_{\rm tx}\Gamma_{u,s}(t)}
\tilde{h}_{u,s}(t),
\end{equation}
where $G_{\rm tx}$ and $G_{\rm rx}$ denote the transmit and receive antenna gains, respectively. For every time slot $t$, user $ u\in\mathcal{U}$, and satellite $s\in\mathcal{S}$, $\Gamma_{u,s}(t)$ denotes the large-scale channel gain, defined as $\Gamma_{u,s} (t)=10^{-{PL}_{u,s}(t)/10}$, where ${PL}_{u,s}(t)$ represents the total large-scale propagation loss. Specifically,
\begin{equation}
\label{eq:pathloss}
PL_{u,s}(t) = \overline{\rm PL}_{u,s}(t) + X_{u,s} + A_{u,s},
\end{equation}
where $\overline{\rm PL}_{u,s}(t) \triangleq L^{\mathrm{fs}}_{u,s}(t) + \bar{L}^{\mathrm{att}}_{u,s}(t)$ is the deterministic free-space path loss, determined by the propagation distance $L^{\mathrm{fs}}_{u,s}(t)$, and the mean atmospheric attenuation caused by atmospheric cloud and rain $\bar{L}^{\mathrm{att}}_{u,s}(t)$. The random variables $X_{u,s}$ and $A_{u,s}$ model the shadow fading and
the residual attenuation about the mean rain and cloud loss, respectively. These large-scale propagation losses follow the 3GPP NTN and ITU-R channel models recommendations~\cite{iturp676,iturp618,iturp840,3gpp38811}. Accordingly, the randomness of the composite channel originates from the shadow fading component $X_{u,s}$, the residual atmospheric attenuation $A_{u,s}$, and the Rician fading $\tilde h_{u,s}(t)$. Based on this stochastic channel model, the instantaneous received SINR between user $u$ and satellite $s$ at time slot $t$ is given by
\begin{equation}
  \gamma_{u,s} (t)
     \;=\; \frac{p_{u,s} (t)\,\lvert h_{u,s}(t)\rvert^{2}}
                {\beta_{u}+\sigma_n^2},
  \label{eq:sinr}
\end{equation}
where $p_{u,s} (t)$ is the transmit power, and $\beta_{u}$ denotes the aggregate co-channel interference from other satellites, and the receiver noise is modelled as additive white Gaussian noise (AWGN) with variance $\sigma_n^2$. Accordingly, the instantaneous achievable rate on the satellite--user link is
\begin{equation}
    r_{u,s}(t)=\frac{W}{B}\,\log_2\left(1+\gamma_{u,s}(t)\right).
\end{equation}

\subsection{Problem Formulation}
\label{sec:hihaap-problem}
In this section, we formulate the proposed chance-constrained handover optimization problem. The objective is to jointly minimize the transmit power consumption, maximize the number of admitted users, and promote stable user--satellite associations while accounting for the stochastic nature of the wireless channel. Because the achievable rate is a random variable under the stochastic channel model, the QoS requirements are enforced through chance constraints that guarantee the target rate with a prescribed probability. Let $i_{u,s}(t)\in\{0,1\}$ denote the association variable for link $(u,s)$ at time slot $t$, where $i_{u,s}(t)=1$ if user $u$ is associated with satellite $s$, and $i_{u,s}(t)=0$ otherwise, for all $u\in\mathcal{U}$ and $s\in\mathcal{S}$. The optimization objective jointly minimizes the normalized transmit power, maximizes user admission, and favours user--satellite associations with higher preference scores. Accordingly, the objective function at time slot $t$ is 
\begin{equation}
\eta(t)=\sum_{u\in\mathcal{U}}\sum_{s\in\mathcal{S}(t)}
\left(\omega_1\frac{p_{u,s}(t)}{P_s^{\max}}-\omega_2-\Pi_{u,s}(t)\right)i_{u,s}(t),
\end{equation}
where $\omega_1,\omega_2\in[0,1]$ are weighting coefficients that balance the trade-off between transmit power minimization and user admission. The constant reward term $-\omega_2$ encourages admitting more users whenever the power and QoS constraints can be satisfied which provides balance between energy efficiency and user admission. The term $\Pi_{u,s}(t)$ denotes the normalized association preference score, which guides the association decision by capturing the trade-off among link quality, expected visibility duration, and association age. Its formulation is presented in Section~\ref{sec:ghaap}. Let $\mathbf{I}(t)$ and $\mathbf{P}(t)$ denote, respectively, the user--satellite association matrix and the transmit-power allocation matrix at time slot $t$. At the beginning of slot $t$, the association realized at the previous slot, $\mathbf{I}(t-1)$, and the association ages $a_u(t) \,\forall{u\in\mathcal{U}}$ are known to the controller. Because the association preference score is a function of $\mathbf{I}(t-1)$ and of the user--satellite geometry only, $\mathbf{\Pi}(t)$ is a known matrix when the slot-$t$ decision is made. Accordingly, at each handover decision period, risk-aware handover problem is formulated  as
\allowdisplaybreaks
\begin{subequations}\label{eq:P0}
\small
\begin{align}
\min_{\mathbf{I}(t),\,\mathbf{P}(t)} & \quad\eta(t) \label{eq:P0-obj}\\
\text{s.t.}\,\,
& \Pr\!\left[\sum_{s\in\mathcal{S}(t)} r_{u,s}(t) \ge r_u^{\mathrm{th}}
   \sum_{s\in\mathcal{S}(t)} i_{u,s}(t)\right]\ge 1-\epsilon, \forall u\in\mathcal{U}, \label{eq:P0-qos}\\
& \frac{p_{u,s}(t)}{P_s^{\max}} \le i_{u,s}(t), \forall u\in\mathcal{U}, \forall s\in\mathcal{S}(t), \label{eq:P0-pow}\\
& \sum_{u\in\mathcal{U}} \frac{p_{u,s}(t)}{P_s^{\max}}\le 1,\ \forall s\in\mathcal{S}(t), \label{eq:P0-total}\\
& i_{u,s}(t)\le V_{u,s}(t),\ \forall u\in\mathcal{U},\ \forall s\in\mathcal{S}(t), \label{eq:P0-visibility}\\
& \sum_{s\in\mathcal{S}(t)} i_{u,s}(t)\le 1,\ \forall u\in\mathcal{U}, \label{eq:P0-assoc}\\
& \sum_{u\in\mathcal{U}} i_{u,s}(t)\le B,\ \forall s\in\mathcal{S}(t), \label{eq:P0-beam}\\
& i_{u,s}(t)\in\{0,1\},\ \forall u\in\mathcal{U},\ \forall s\in\mathcal{S}(t).\label{eq:P0-bin}
\end{align}
\end{subequations}
The objective in~\eqref{eq:P0-obj} jointly minimizes the transmit power while maximizing the admission of users through the association variables and favouring user--satellite links with higher association preference scores. Constraint~(7b) is a probabilistic QoS constraint that ensures the minimum rate requirement is satisfied with probability at least $1-\epsilon$, accounting for channel fading. Therefore, it accounts for the stochastic nature of the wireless channel. Constraint~\eqref{eq:P0-pow} ensures that transmit power is only allocated to users associated with the satellite by coupling the power allocation and association variables. Constraint~\eqref{eq:P0-total} ensures that the total transmit power allocated to all associated users does not exceed the maximum transmit power budget of each satellite. To avoid infeasible associations,~\eqref{eq:P0-visibility} guarantees that a user can only be associated with satellites that are visible at the corresponding time slot. Constraint~\eqref{eq:P0-assoc} ensures that each user is associated with at most one satellite at every time slot. Constraint~\eqref{eq:P0-beam} limits the number of users simultaneously served by each satellite according to its available beams. Finally,~\eqref{eq:P0-bin} enforces the binary nature of the association variables.

\subsection{GHAAP Score} \label{sec:ghaap}
The association decision jointly accounts for the link quality, expected remaining visibility duration, and association age to balance communication performance and handover stability. In other words, relying only on the link quality may result in frequent handovers, whereas considering only the remaining visibility duration may lead to associations with poor communication quality. Therefore, an effective association metric should jointly capture both communication performance and connection stability. To this end, we introduce the GHAAP score that accounts for all these factors. The normalized GHAAP score, $\Pi_{u,s}(t)\in[0,1]$, for all $u\in\mathcal{U}$ and $s\in\mathcal{S}$, is defined as

\begin{equation} 
\Pi_{u,s}(t) = \frac{\tilde{\Pi}_{u,s}(t)} {\max_{s' \in \mathcal{S}_u(t)} \tilde{\Pi}_{u,s'}(t)}. 
\end{equation}
where,
\begin{equation}
\label{eq:pref}
\tilde{\Pi}_{u,s}(t)
\triangleq
\hat{\tau}_{u,s}^{\,\alpha_1}(t)\hat{g}_{u,s}^{\,1-\alpha_1}(t)
+
\frac{\alpha_2 T^{\mathrm{pass}}\,i_{u,s}(t-1)}
{\alpha_2 T^{\mathrm{pass}}+(1-\alpha_2)a_u(t)}.
\end{equation}
A larger $\Pi_{u,s}(t)$ favours associations that balance communication quality and connection stability. We introduce $\widetilde{\Pi}_{u,s}(t)$ as an association-preference metric, where $\alpha_1,\alpha_2\in[0,1]$ are design parameters. Specifically, $\alpha_1$ controls the tradeoff between the predicted remaining visibility duration and link quality, while $\alpha_2$ manage the contribution of association persistence. These parameters can be selected empirically through a parameter sweep according to the decision maker's objectives. Here,
$
\hat{\tau}_{u,s}(t)
=
\frac{\tau_{u,s}(t)}
{\displaystyle\max_{s'}\tau_{u,s'}(t)}
$
denotes the normalized remaining visibility duration, while
$
\hat{g}_{u,s}(t) =
\frac{\log_2\!\left(1+\left(d_{\max,u}(t)/d_{u,s}(t)\right)^{2}\right)}
     {\max_{s' \in \mathcal{S}_u(t)}
      \log_2\!\left(1+\left(d_{\max,u}(t)/d_{u,s'}(t)\right)^{2}\right)}
$
is the distance-based link-quality metric based on the user--satellite distance, where $d_{u,s}(t)$ denotes the instantaneous Euclidean distance between user $u$ and satellite $s$ at time slot $t$, and
$
d_{\max,u}(t)=\max_{s'}d_{u,s'}(t)
$
is the maximum distance between user $u$ and the visible satellites in $\mathcal{S}_u(t)$. Moreover, $\tau_{u,s}(t)$ denotes the predicted remaining visibility duration between user $u$ and satellite $s$ at time slot $t$, $a_u(t)$ denotes the age of the current association, i.e., the elapsed time since the last handover, and $T^{\mathrm{pass}}$ is the persistence time constant. Both $\hat{\tau}_{u,s}(t)$ and $\hat{g}_{u,s}(t)$ are normalized by their maxima over $s' \in \mathcal{S}_u(t)$ and hence lie in $(0,1]$, so their geometric combination is scale-free.

The GHAAP score consists of two components. The first part, referred to as the {link preference}, evaluates each link according to its expected remaining visibility duration and communication quality. The weighting parameter $\alpha_1\in[0,1]$ controls the trade-off between these two factors. Larger values give greater preference to the remaining visibility duration, while smaller values prioritize link quality. The latter, referred to as the {association persistence}, introduces a bias toward the currently serving satellite to discourage unnecessary handovers. This persistence term is activated only for the currently serving satellite. The parameter $\alpha_2\in[0,1]$ controls the persistence mechanism, with higher values assigning more preference to maintain the current association

\section{Solution Approach}\label{sec:solution}
Problem~\eqref{eq:P0} is a non-convex MINLP, making it generally NP-hard. To obtain a computationally tractable solution, the problem is decomposed into two subproblems, namely, an inner power allocation subproblem and an outer association subproblem. The inner subproblem determines the minimum transmit power required to satisfy the QoS constraint. The outer subproblem optimizes the association variables based on the optimal transmit power obtained from the inner subproblem while satisfying the visibility and resource constraints.\vspace{-.1 in}

\subsection{Problem Decomposition}
The proposed solution is motivated by the observation that, for any fixed set of users and satellites, the transmit-power allocation problem is separable in the level of the admitted user--satellite links. Moreover, the objective function is increasing in the transmit powers, so any feasible solution allocating transmit power above the minimum required to satisfy the corresponding chance constraint is suboptimal. Consequently, the optimal transmit power for every admitted link is the minimum value satisfying its chance constraint with equality, i.e., the corresponding chance constraint is active at the optimum, and the optimal power is obtained directly in closed form. Substituting these optimal transmit powers into~\eqref{eq:P0} eliminates the continuous power variables and yields an equivalent optimization problem over the association variables.
\subsubsection{Power Allocation Subproblem}
 For each user--satellite link $(u,s)$ assuming $i_{u,s}(t)=1$, the inner subproblem determines the minimum transmit power required to satisfy the QoS requirement while respecting the maximum transmit power constraint. Accordingly, the link-level power-allocation problem is formulated as
\begin{subequations}\label{eq:P1}
\begin{align}
\min_{p_{u,s}(t)}
\quad
&
p_{u,s}(t)
\label{eq:P1-obj}
\\
\text{s.t.}\quad\;
&
\Pr\!\left[
\frac{p_{u,s}(t)|h_{u,s}(t)|^{2}}
{\beta_u+\sigma_n^{2}}
\ge
\gamma_u^{\rm th}
\right]
\ge
1-\epsilon,
\label{eq:P1-qos}
\\
&
p_{u,s}(t)\le P_s^{\max},
\label{eq:P1-pow}
\end{align}
\end{subequations}
where $\gamma_u^{\rm th}=2^{r_u^{\rm th}/(W/B)}-1$ is the minimum SINR required to satisfy the target rate requirement $r_u^{\rm th}$. The channel power gain in \eqref{eq:channel} can be expressed as
\begin{equation}
|h_{u,s}(t)|^{2} = \bar g_{u,s}(t) Z_{u,s}(t), \label{eq:channelpower}
\end{equation}
where $\bar{g}_{u,s}(t) = G_{\mathrm{tx}} G_{\mathrm{rx}} 10^{-\overline{\rm PL}_{u,s}(t)/10}$ denotes the deterministic component of the channel gain and $Z_{u,s}(t) = 10^{-(X_{u,s}+A_{u,s})/10} |\tilde{h}_{u,s}(t)|^2$ is the composite fading coefficient. The shadow fading $X_{u,s}$ and the residual rain and cloud attenuation $A_{u,s}$are each modelled as independent Gaussian random variables in dB~\cite{10153617,10622620}. Then, $X_{u,s}+A_{u,s}\sim\mathcal{N}(0,\sigma_{\mathrm{tot}}^{2})$, where $\sigma_{\mathrm{tot}}=\sqrt{\sigma^{2}+\sigma_{a}^{2}}$~\cite{10153617,10622620}, while $|\tilde{h}{u,s}(t)|^{2}$ denotes the unit-mean Rician fading power with the elevation-dependent Rician factor $K{u,s}(t)$. The composite lognormal--Rice random variable $Z_{u,s}(t)$ does not admit a tractable closed-form cumulative distribution function. Therefore, we adopt the equivalent-lognormal approximation of~\cite{4275022} to derive a closed-form solution according to following proposition.


\begin{proposition}
\label{prop:minpower}
The minimum transmit power satisfying the chance constraint
\eqref{eq:P1-qos} is
\begin{equation}
p_{u,s}^{\min}(t,\epsilon)
=
\frac{
\gamma_u^{\rm th}
(\beta_u+\sigma_n^2)
}
{\bar g_{u,s}(t)}
10^{D(\sigma_{\rm tot},K_{u,s}(t),\epsilon)/10},
\label{eq:pmin}
\end{equation}
where
$
D(\sigma_{\rm tot},K_{u,s}(t),\epsilon)
=
-\left(
\mu_Y+\sigma_Y\Phi^{-1}(\epsilon)
\right),
$
and $(\mu_Y,\sigma_Y)$ denote the logarithmic parameters of the equivalent lognormal random variable. $\Phi^{-1}(\cdot)$ denotes the inverse cumulative distribution function of the standard normal distribution.
\end{proposition}
For brevity, the detailed proof is provided in Appendix~\ref{app:mgf}. The proposition illustrates that the equivalent-lognormal margin $D(\sigma_{\rm tot},K,\epsilon)$ depends only on the composite shadowing standard deviation $\sigma_{\rm tot}$, the Rician factor $K$, and the target outage probability $\epsilon$. Therefore, it is precomputed offline by solving the moment generating function (MGF) matching equations of over a discretized $(\sigma_{\rm tot},K,\epsilon)$ parameter space and stored as a lookup table. During runtime, the proposed algorithm retrieves the corresponding reserve margin through a constant-time lookup and computes the minimum transmit power using \eqref{eq:pmin}, thereby avoiding repeated numerical solutions of the MGF-matching equations while preserving the equivalent-lognormal approximation.

\subsubsection{Outer Association Subproblem}

Substituting the optimal transmit power $p_{u,s}^{\min}(t,\epsilon)$ obtained from Proposition~\ref{prop:minpower} into the original optimization problem eliminates the power allocation variables while guaranteeing satisfaction of the QoS constraint. The original MINLP reduces to the following association optimization problem:
\begin{subequations}\label{eq:P2}
\begin{align}
\min_{\mathbf{I}(t)}\;\;
& \sum_{u\in\mathcal{U}}\sum_{s\in\mathcal{S}(t)}
\left(
\omega_1\frac{p_{u,s}^{\min}(t,\epsilon)}{P_s^{\max}}
-\omega_2-\Pi_{u,s}(t)
\right)i_{u,s}(t)
\label{eq:P2-obj}\\
\text{s.t.}\quad
& \sum_{u\in\mathcal{U}} p_{u,s}^{\min}(t,\epsilon)i_{u,s}(t)\le P_s^{\max},
\quad \forall s\in\mathcal{S}(t),
\label{eq:P2-total}\\
& i_{u,s}(t)\le V_{u,s}(t),
\quad \forall u\in\mathcal{U},~\forall s\in\mathcal{S}(t),
\label{eq:P2-visibility}\\
& \sum_{s\in\mathcal{S}(t)} i_{u,s}(t)\le1,
\quad \forall u\in\mathcal{U},
\label{eq:P2-assoc}\\
& \sum_{u\in\mathcal{U}} i_{u,s}(t)\le B,
\quad \forall s\in\mathcal{S}(t),
\label{eq:P2-beam}\\
& i_{u,s}(t)\in\{0,1\},
\quad \forall u\in\mathcal{U},~\forall s\in\mathcal{S}(t).
\label{eq:P2-bin}
\end{align}
\end{subequations}
The resulting association subproblem is a integer ILP and equivalently a generalized assignment problem (GAP)~\cite{maniezzo2021generalized}, in which users are assigned to satellites while satisfying the satellite power budget and beam-capacity constraints. The problem can be solved to optimality using commercial mixed-integer optimization solvers for moderate problem sizes, such as \texttt{MOSEK} \cite{mosek}.

\subsection{Proposed Solution Algorithms}
 Algorithm~\ref{alg:optimal} illustrates the proposed solution. The algorithm begins by solving the power allocation subproblem to obtain the minimum required transmit power, $p_{u,s}^{\rm min}(t,\epsilon)$, for each visible user--satellite link. Let $\mathcal{F}$ denote the set of feasible user--satellite pairs and $\mathcal{U}_{\mathrm{f}}$ the set of users having at least one feasible connection satisfying the power and visibility constraints. The set of feasible pairs is then fed to the ILP, and the corresponding minimum-power values are substituted into the association objective, resulting in an ILP that jointly minimizes the normalized transmit power cost while maximizing user admission through the association preference metric. The final power allocation is obtained as $\mathbf{P}(t)=\mathbf{I}(t)\odot\mathbf{P}^{\min}(t,\epsilon)$, where $\odot$ denotes the element-wise product. 

To further reduce the computational complexity of the ILP solution in Algorithm~\ref{alg:optimal}, we propose the greedy two-stage algorithm presented in Algorithm~\ref{alg:proposed}. The algorithm starts by solving the power allocation subproblem \eqref{eq:P1} to compute the minimum required transmit power, $p_{u,s}^{\rm min}(t,\epsilon)$, for each visible user--satellite link while satisfying the target QoS outage probability.  Subsequently, infeasible user--satellite pairs requiring more than the available satellite power are discarded. Then, the association cost of each feasible link is computed according to the objective coefficients of~\eqref{eq:P2} as $c_{u,s}(t)=\omega_1\frac{p_{u,s}^{\rm min}(t,\epsilon)}{P_s^{\max}}-\omega_2-\Pi_{u,s}(t)$. Next, we solve a relaxed association subproblem by removing the total power-budget constraint. The resulting problem is formulated as

\allowdisplaybreaks
\begin{subequations}
\label{eq:LAP1}
\begin{align}
\min_{\mathbf{I}(t)}\quad
&\sum_{u\in\mathcal{U}_{\mathrm{f}}}\sum_{s\in\mathcal{S}(t)}
c_{u,s}(t)i_{u,s}(t)\\
\text{s.t.}\quad
&\sum_{s\in\mathcal{S}(t)}i_{u,s}(t)=1,\quad
\forall u\in\mathcal{U}_{\mathrm{f}},\\
&\sum_{u\in\mathcal{U}_{\mathrm{f}}}i_{u,s}(t)\le B,\quad
\forall s\in\mathcal{S}(t),\label{cons:14c}\\
&i_{u,s}(t)\in\{0,1\},\;
i_{u,s}(t)=0,\quad
\forall (u,s)\notin\mathcal{F}.
\end{align}
\end{subequations}
Problem~\eqref{eq:LAP1} is transformed into a standard linear assignment problem by representing each satellite with beam capacity $B$ as $B$ virtual satellites with unit capacity. Consequently, the capacity constraint~\eqref{cons:14c} is replaced by a one-to-one assignment over the expanded set of virtual satellites. The resulting problem can therefore be solved optimally using the Hungarian algorithm with polynomial-time complexity. If the resulting association violates the satellite power budget, each overloaded satellite solves a low-complexity 0--1 knapsack problem to determine the optimal set of users to  reassigned.
\begin{subequations}
\label{eq:KNAP}
\begin{align}
\min_{I_{s}(t)}\quad
&\sum_{u\in\mathcal{U}_{s}(t)} c_{u,s}(t)i_{u,s}(t),\\
\text{s.t.}\quad
&\sum_{u\in\mathcal{U}_{s}(t)} w_{u,s}(t)i_{u,s}(t)\le 1,
\end{align}
\end{subequations}
where $\mathcal{U}_{s}(t)=\{u:i_{u,s}(t)=1\}$ denotes the set of users assigned to satellite $s$ at time slot $t$, and $
w_{u,s}(t)={p_{u,s}^{\rm min}(t,\epsilon)}/{P_s^{\max}}
$
is the normalized fraction of the satellite power budget required for user $u$. Because each satellite supports at most $B$ users, i.e., $|\mathcal{U}_{s}(t)|\le B$,~\eqref{eq:KNAP} can be solved exactly by evaluating all $2^{\rm B}$ feasible user subsets, resulting in a worst-case complexity of $\mathcal{O}(2^{\rm B})$, or approximately using an linear programming (LP) relaxation of the binary variables, followed by a rounding procedure to obtain the binary solution. The knapsack problems for all overloaded satellites are solved in parallel. The users excluded from the optimal subsets $\mathcal{U}_{s}^{\star}$ are then sequentially considered for reassignment to feasible satellites with sufficient beam and power capacity. After each successful reassignment, the corresponding satellite load is immediately updated to guarantee the beam-capacity and power-budget constraints. 

\begin{algorithm}[!t]
\caption{ILP-Based Risk-Aware Association}
\label{alg:optimal}
\begin{algorithmic}[1]
\Require $\mathcal{U}$, $\mathcal{S}(t)$, $V_{u,s}(t)$, $P_s^{\max}$, $B$, $\epsilon$, $\gamma_u^{\rm th}$, $z_{1-\epsilon}\!=\!\Phi^{-1}(1-\epsilon)$
\Ensure $\mathbf{I}(t)$, $\mathbf{P}(t)$

\ForAll{$(u,s)$ with $V_{u,s}(t)=1$}
    \State Compute $p_{u,s}^{\min}(t,\epsilon)$ using \eqref{eq:pmin}
\EndFor

\State $\mathcal{F}\gets\{(u,s):V_{u,s}(t)=1,\,
p_{u,s}^{\min}(t,\epsilon)\le P_s^{\max}\}$

\State
$\mathcal{U}_{\mathrm f}(t)\gets
\left\{
u:\exists s,\;
(u,s)\in\mathcal{F}(t)
\right\}$

\ForAll{$(u,s)\in\mathcal{F}$}
    \State compute $c_{u,s}$ 
\EndFor

\State $\mathbf{I}(t)\gets$
\Call{ILP}{$c,\mathcal{F},B,P_s^{\max}$}
\Comment{Solve \eqref{eq:P2}}

\State $\mathbf{P}(t) \gets \mathbf{I}(t)\odot\mathbf{P}^{\min}(t,\epsilon)$

\State \Return $\mathbf{I}(t),\mathbf{P}(t)$
\end{algorithmic}
\end{algorithm}

\subsection{Computational and Convergence Analysis Complexity}
The minimum required transmit powers and the corresponding association costs are computed in closed form with complexity $\mathcal{O}(\rm S\, \rm U)$. The relaxed association problem is then solved optimally using the Hungarian algorithm with complexity $\mathcal{O}({\rm U}^{2}\,\rm B\,\rm S)$, which dominates the overall computational cost. The budget repairing stage solves an LP relaxation of the knapsack problem for each overloaded satellite, followed by binary rounding and feasibility recovery step. Because each satellite serves at most \(\rm B\) users, these LPs have at most \(\rm B\) variables. Therefore, the overall complexity of the proposed algorithm is $\mathcal{O}({\rm U}^{2}\,\rm B\,\rm S)$. For a fixed $B$, the complexity is dominated by the Hungarian assignment stage and is polynomial in $|\mathcal{U}|$ and $|\mathcal{S}|$. In contrast, the optimal ILP formulation in~\eqref{eq:P2} retains the aggregate satellite power-budget constraint and requires exponential-time worst-case complexity in the number of binary association variables, $\rm S \rm U$, when solved exactly using branch-and-bound. Hence, for a fixed $\rm B$, the proposed algorithm reduces the worst-case scaling with respect to $\rm U $ and $\rm S $ from exponential to polynomial. Algorithm~\ref{alg:proposed} does not require an iterative convergence procedure. Each overloaded satellite is repaired once using the knapsack problem. This is followed by sequential reassignment of the excluded users subject to the beam-capacity and power-budget constraints. Since the satellite load is updated after each reassignment, no new power-constraint violation is introduced. Therefore, the algorithm terminates after one repair round and a finite reassignment pass with a feasible association.%
\begin{algorithm}[!t]
\caption{Greedy Two-Stage Risk-Aware Association}
\label{alg:proposed}
\begin{algorithmic}[1]
\Require $\mathcal{U}$, $\mathcal{S}(t)$, $V_{u,s}(t)$, $P_s^{\max}$, $B$,
$\epsilon$, $\gamma_u^{\rm th}$,
$z_{1-\epsilon}\gets\Phi^{-1}(1-\epsilon)$
\Ensure $\mathbf{I}(t),\mathbf{P}(t)$

\ForAll{$(u,s)$ with $V_{u,s}(t)=1$}
    \State Compute $p_{u,s}^{\min}(t,\epsilon)$ using \eqref{eq:pmin}
\EndFor

\State
$\mathcal{F}(t)\gets
\left\{
(u,s):
V_{u,s}(t)=1,\;
p_{u,s}^{\rm min}(t,\epsilon)\le P_s^{\max}
\right\}$

\State
$\mathcal{U}_{\mathrm f}(t)\gets
\left\{
u:\exists s,\;
(u,s)\in\mathcal{F}(t)
\right\}$

\ForAll{$(u,s)\in\mathcal{F}(t)$}
    \State Compute $w_{u,s}(t)$ and $c_{u,s}(t)$
\EndFor

\State
$\mathbf I(t)\gets
\Call{Hungarian}
{\big(c(t),\mathcal{F}(t),\mathcal U_{\mathrm f}(t),B\big)}$

\ForAll{$s\in\mathcal S(t)$}
    \State
    $\ell_s\gets
    \sum_u w_{u,s}(t)i_{u,s}(t)$
    \State
    $n_s\gets
    \sum_u i_{u,s}(t)$
\EndFor

\If{$\ell_s\le1,\;\forall s\in\mathcal S(t)$}
    \State
    $\mathbf P(t)\gets
    \mathbf I(t)\odot\mathbf P^{\min}(t,\epsilon)$
    \State
    \Return $\mathbf I(t),\mathbf P(t)$
\EndIf

\State
$\mathcal S_{\rm ov}(t)\gets
\{s\in\mathcal S(t):\ell_s>1\}$

\ForAll{$s\in\mathcal S_{\rm ov}(t)$ \textbf{in parallel}}

    \State
    $\mathcal U_s(t)\gets
    \{u:i_{u,s}(t)=1\}$

    \State
    $\mathcal U_s^\star\gets
    \Call{Knapsack}
    {\mathcal U_s(t),
    c_{\cdot,s}(t),
    w_{\cdot,s}(t),
    1}$

    \ForAll{$u\in
    \mathcal U_s(t)\setminus\mathcal U_s^\star$}
        \State
        $i_{u,s}(t)\gets0$
    \EndFor

\EndFor

\State
Update $\ell_s$ and $n_s$, $\forall s\in\mathcal S(t)$

\ForAll{$u\in\mathcal U_{\mathrm f}(t)$ with
$\sum_s i_{u,s}(t)=0$}

    \State
    $\mathcal C\gets
    \left\{
    s':
    (u,s')\in\mathcal F(t),\;
    n_{s'}<B,\;
    \ell_{s'}+w_{u,s'}(t)\le1
    \right\}$

    \If{$\mathcal C\neq\emptyset$}

        \State
        $\hat s\gets
        \arg\max_{s'\in\mathcal C}
        \Pi_{u,s'}(t)$

        \State
        $i_{u,\hat s}(t)\gets1$, $\ell_{\hat s}\gets
        \ell_{\hat s}+w_{u,\hat s}(t)$,
        $n_{\hat s}\gets
        n_{\hat s}+1$

    \EndIf

\EndFor

\State
$\mathbf P(t)\gets
\mathbf I(t)\odot\mathbf P^{\min}(t,\epsilon)$

\State
\Return $\mathbf I(t),\mathbf P(t)$

\end{algorithmic}
\end{algorithm}

\section{ Simulation Results}
\label{sec:results}
In this section, we validate the performance of the proposed handover management framework through numerical experiments using the proposed \texttt{NTN-Connect} simulator. The experiments are conducted on the OneWeb and Starlink constellations, which differ in satellite density and latitude coverage. To do so, we first introduce the \texttt{NTN-Connect} simulator. Then, we describe the simulation setup, followed by the performance of the proposed algorithms. Finally, we present its performance of the proposed framework against the baselines.

\subsection{Simulator Framework}
\begin{figure}[t]
\centering
\includegraphics[width=.96\columnwidth,draft=false]{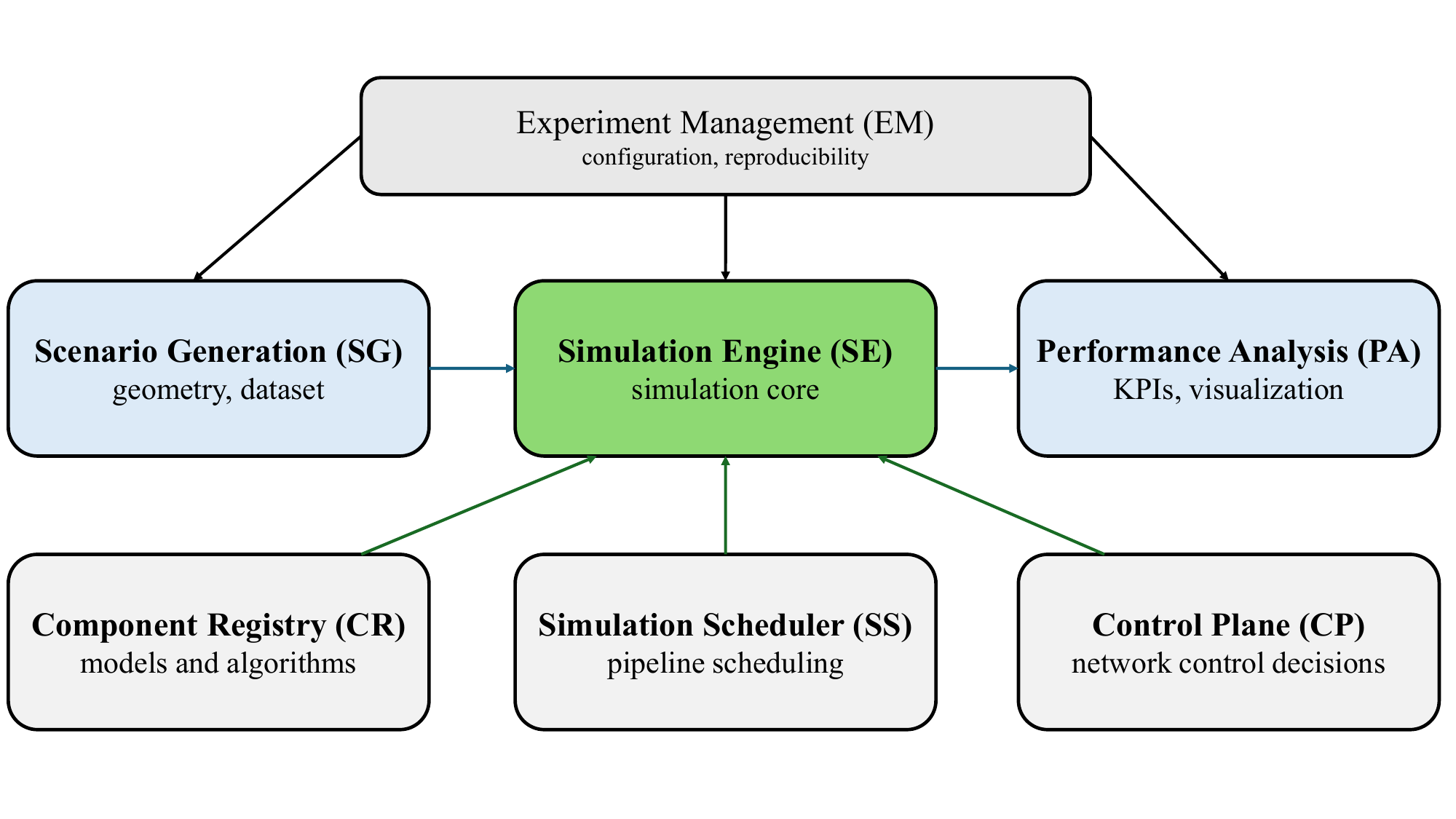}
\caption{\texttt{NTN-Connect} architecture.}
\label{fig:architecture}
\vspace{-.09in}
\end{figure}

\texttt{NTN-Connect} is a modular framework for system-level simulation of  NTNs. Fig.~\ref{fig:architecture} presents an overview of the \texttt{NTN-Connect} architecture and the data flow between its entities. Execution follows a three-stage workflow comprising the scenario generation (SG), the simulation engin (SE), and the performance analysis (PA). SG constructs the network environment and exports it as a reusable dataset. The SE evaluates communication and networking over the generated dataset under the scheduling of the simulation scheduler (SS). Finally, the PA reports the resulting key performance indicators (KPIs). Four supporting components operate across the framework. The component registry (CR) provides interchangeable models and algorithms. Experiment management (EM) manages configuration and reproducible execution. The simulation scheduler (SS) schedules the execution of the simulation pipeline. Finally, the control plane (CP) decouples network decision making from simulation execution while supporting both centralized and distributed operation. \texttt{NTN-Connect} follows three design principles: (i) modularity, (ii) integration, and (iii)programmability, and reproducibility, which enable configurable components, cross-layer interactions through a shared network state, and reproducible execution.

\subsubsection{Scenario Generation (SG)}
\label{sec:architecture:scenario}
SG is the first stage of the \texttt{NTN-Connect} workflow. It constructs the time-varying network environment, mainly the geometry and visibility data that define the structure of the NTN. \texttt{NTN-Connect} supports analytical satellite generation using Walker constellations \cite{walker1984satellite}. Alternatively, constellations can be imported from two-line element (TLE) data. The terrestrial nodes, including the gateways and base stations, are defined by their geographic locations, while the user terminals are placed according to spatial-distribution models, including uniform-random and Fibonacci-lattice sampling \cite{gonzalez2010fibonacci}. Meanwhile, satellites are propagated using the SGP4/SDP4 models through the \texttt{Skyfield} library\cite{rhodes2019skyfield}. The propagator is itself a registered component, so other models, a closed-form Keplerian propagator~\cite{vallado2013}, can be added to the registry. To limit the cost of mega-constellation studies, satellites are propagated in two passes: a coarse pass discards satellites never visible to any ground node, and a fine pass propagates only the retained set.

\subsubsection{Simulation Engine (SE)}
The SE is the computational core of \texttt{NTN-Connect}. Using the dataset files generated by SG, the SE evaluates the physical-, link-, and network-layer procedures of the simulation. These procedures are scheduled by the SS, which advances the simulation in discrete time steps and schedules the component pipeline once per interval over a shared simulation state. This state holds a snapshot of the network at each interval, including user and satellite states, link and network conditions, and the performance metrics the controllers use for decision-making. Accordingly, the SE executes a sequence of functional components in the order the SS schedules. These components include mobility models, channel models, handover and routing algorithms, RRM, and metric collection. Mobility models describe the movement of network entities, while channel models generate wireless channel realizations; the remaining components perform association, routing, resource allocation, and metric collection. Their modular design enables each component to be independently extended without affecting the framework.
\begin{table}[t]
\centering
\caption{Simulation parameters.}
\label{table:parameters}
\small
\renewcommand{\arraystretch}{1}
\begin{tabular}{|l|c|l|}
\hline
\textbf{Parameter} & \textbf{Symbol} & \textbf{Value} \\
\hline\hline
Region                       & --                 & Canada (Quebec) \\
\hline
Number of ground users       & $\rm U$       & $50$ \\
\hline
OneWeb satellites       & $\rm S$       & $649$ \\
\hline
Starlink satellites    & $\rm S$       & $1{,}343$ \\
\hline
Min.\ elevation (user)       & $\theta_{\rm min}^u$    & $25^\circ$ \\
\hline
Simulation duration          & $T$                & $45$~min \\
\hline
Handover decision period     & $T_{\rm ho}$       & $2$~s \\
\hline
Carrier frequency            & $f_c$              & $12$~GHz \\
\hline
System bandwidth  (per satellite)  & $W$                & $500$~MHz \\
\hline
Satellite EIRP (OneWeb)      & $\mathrm{EIRP}$    & $58.5$~dBW \\
\hline
Satellite EIRP (Starlink)    & $\mathrm{EIRP}$    & $52$~dBW \\
\hline
User terminal figure of merit& $G/T$              & $-5.8$~dB/K \\
\hline
Per-satellite power budget   & $P_s^{\rm max}$    & $20$~dBW \\
\hline
Target spectral efficiency & $r_u^{\mathrm{th}}B/W$ & $1$--$5$ bps/Hz \\
\hline
Beams per satellite          & $\rm B$            & $8$ \\
\hline
Outage probability           & $\epsilon$         & $0.05$ \\
\hline
Power weight                     & $\omega_1$             & $1$ \\
\hline
Admission reward                 & $\omega_2$             & $0.25$ \\
\hline
Pass constant (Starlink,OneWeb)         & $T^{\mathrm{pass}}$    & $(180, 300)$ s \\
\hline
\end{tabular}
\vspace{-.0 in}
\end{table}
\begin{figure*}[t!]
\centering
\begin{subfigure}[b]{0.32\textwidth}
  \centering
  \includegraphics[width=\linewidth,draft=false]{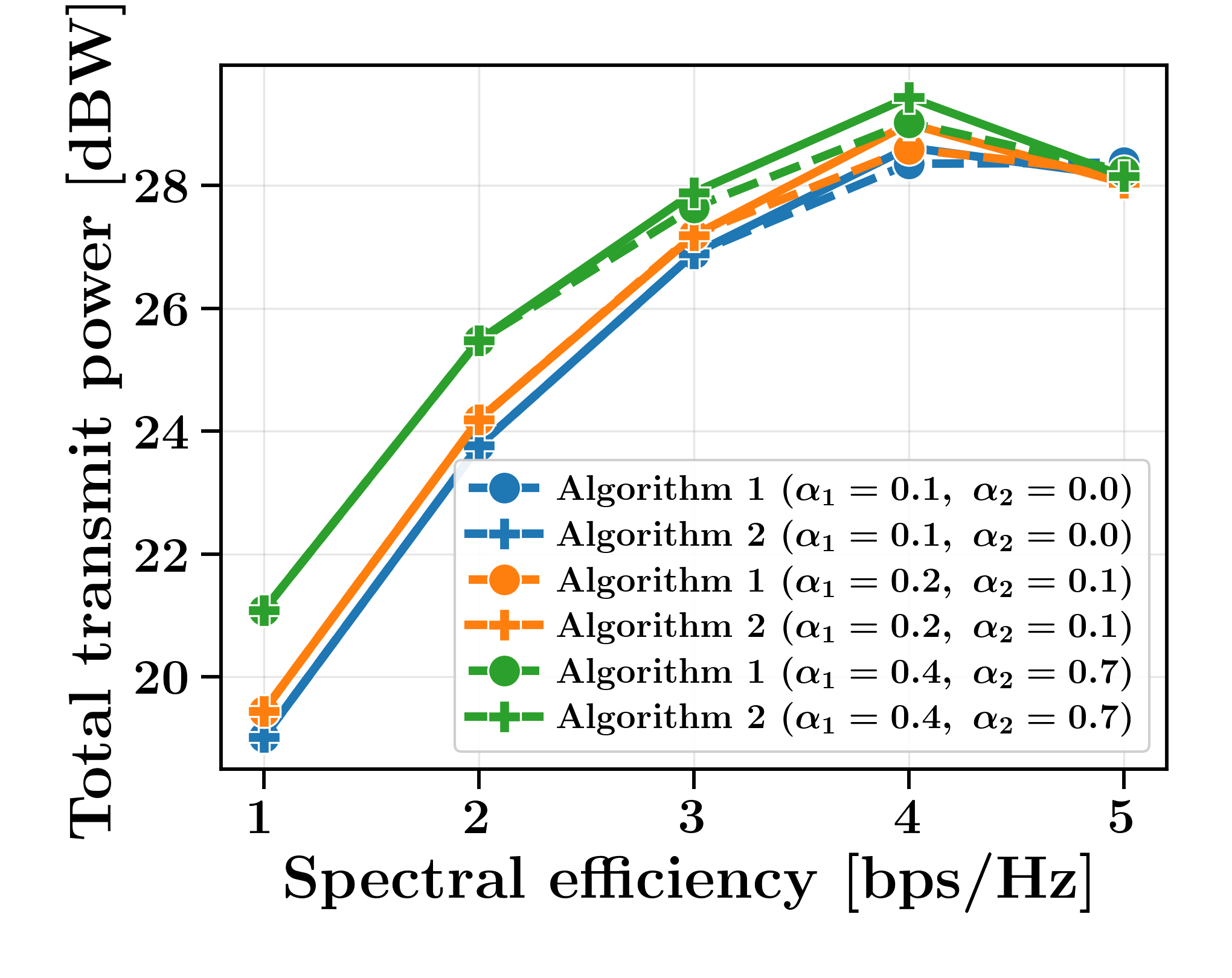}
  \caption{Transmit power [dBW]}
  \label{Fig:opt_ho_power_oneweb}
\end{subfigure}\hfill
\begin{subfigure}[b]{0.32\textwidth}
  \centering
  \includegraphics[width=\linewidth,draft=false]{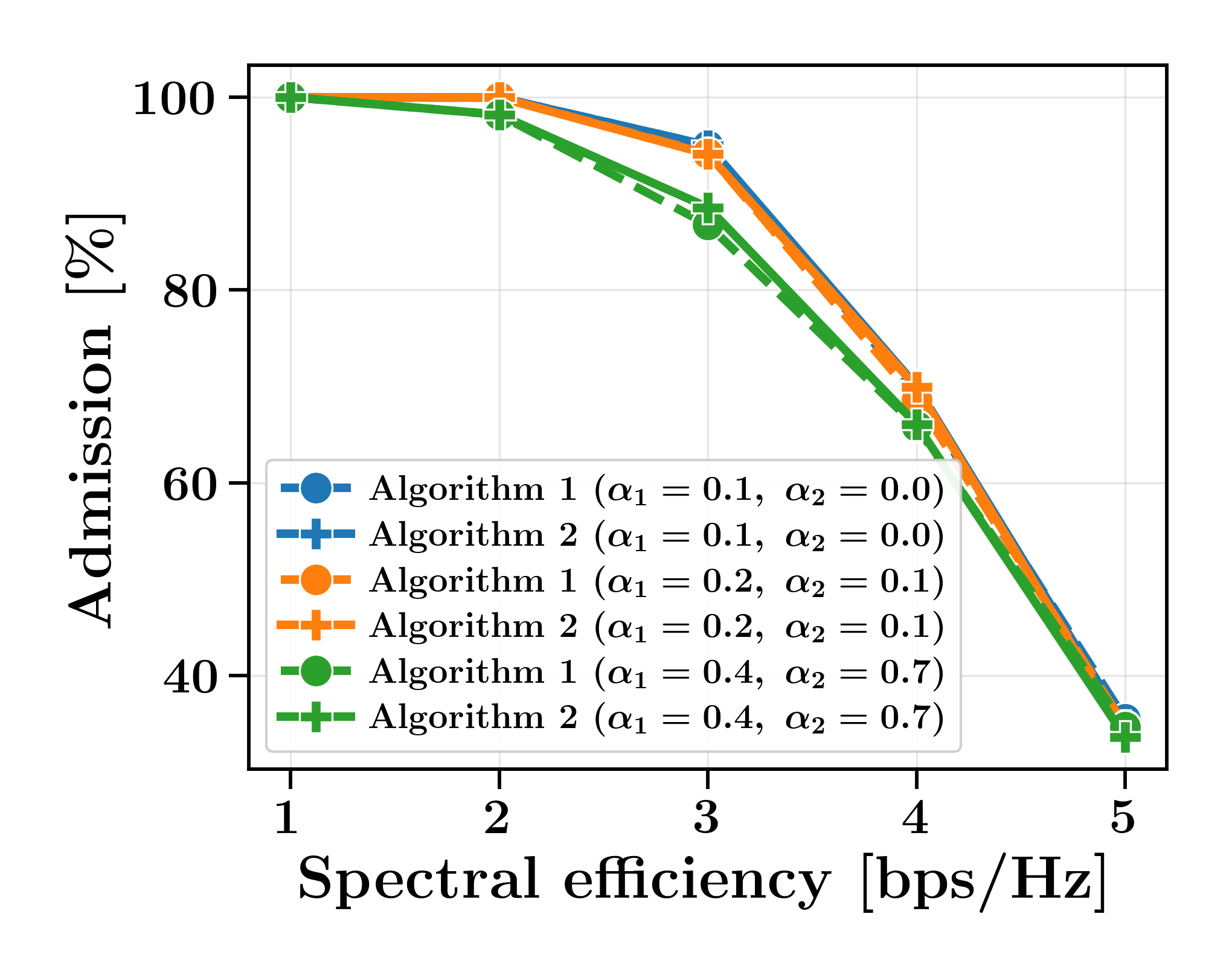}
  \caption{Admission}
  \label{Fig:opt_adm_oneweb}
\end{subfigure}\hfill
\begin{subfigure}[b]{0.32\textwidth}
  \centering
  \includegraphics[width=\linewidth,draft=false]{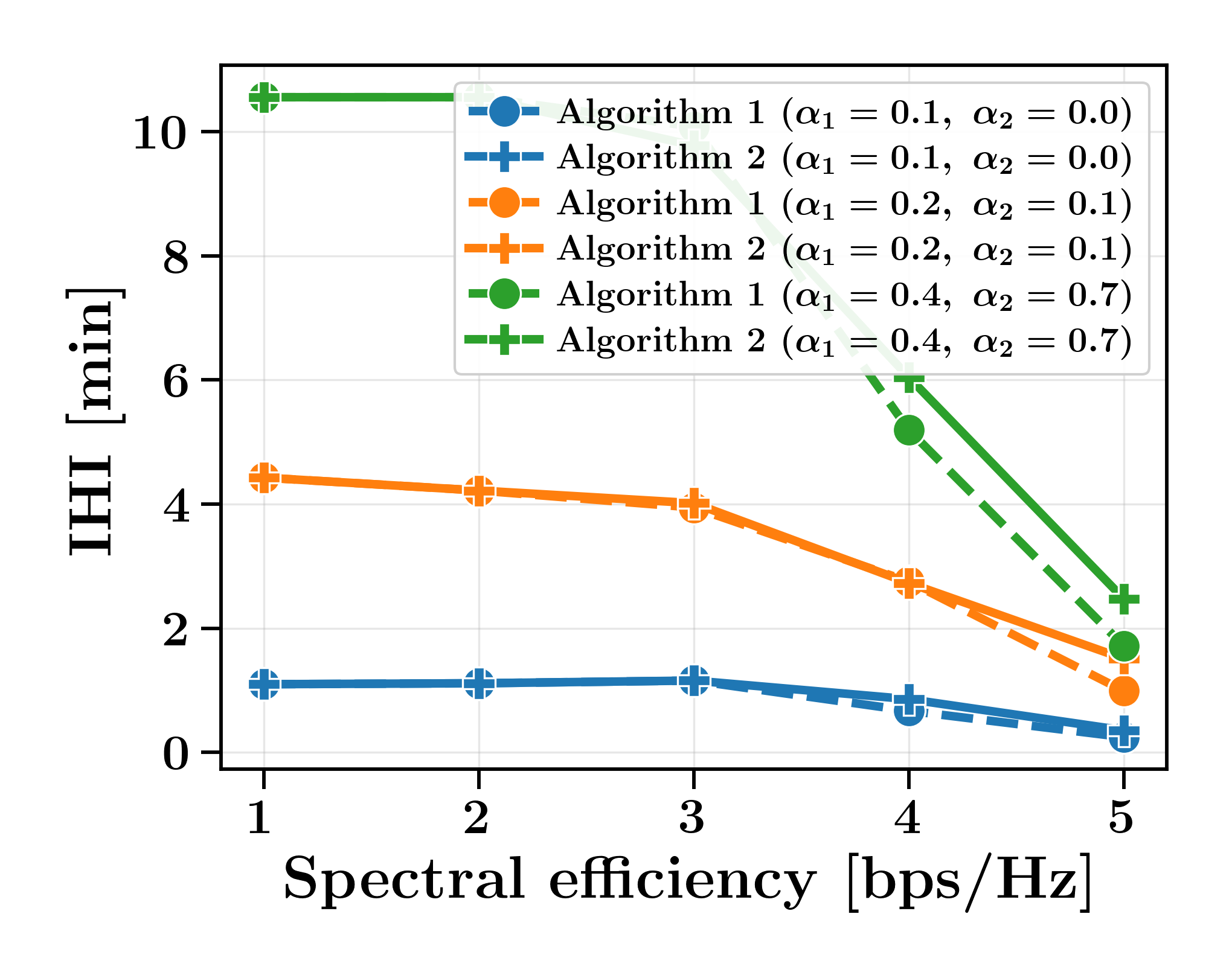}
  \caption{IHI}
  \label{Fig:opt_ihi_oneweb}
\end{subfigure}
\caption{Performance comparison of the proposed Algorithm 1 and Algorithm 2.}
\label{fig:optimality}
\vspace{-.2 in}
\end{figure*}
\subsubsection{Performance Analysis (PA)}
\label{sec:architecture:analysis}
PA is the final stage of the \texttt{NTN-Connect} workflow. It translates the results produced during the simulation into the corresponding KPIs. These KPIs include link-performance metrics, such as spectral efficiency and achievable rate; network-level metrics, such as E2E latency and routing statistics; as well as mobility-management and resource-utilization metrics. It presents them through multiple statistical representations, including cumulative and complementary cumulative distribution functions (CDFs and CCDFs), summary statistics, and confidence intervals.
\subsubsection{Support Services}
The framework is supported by four components, including SS, CR, CP, and EM. The SS coordinates the execution of the SE by scheduling the component pipeline while maintaining a shared simulation state. The SS validates the execution order and detects conflicting state updates. The CR separates the simulation infrastructure from the algorithmic components to enable modularity. This helps to integrate models and algorithms through configuration rather than modifying the SE. The CR organizes functionality into interchangeable components with common interfaces using registry-based plugins. During the simulation, the CP governs network decision-making by translating the shared simulation state into control actions, such as user association. The CP consists of programmable controllers that read the shared state, execute the configured decision or optimization algorithm, and write the resulting actions before the next pipeline stage. EM manages experiment specification, execution, and result recording to ensure reproducible simulations. EM configures simulation scenarios through configuration files, initializes deterministic random streams from a master seed, and records configuration, metadata, execution statistics, and performance metrics for each experiment.\vspace{-.1 in}

\subsection{Simulation Setup}
\label{sec:setup}

The analysis is performed over a Canadian service region, where satellites serve a set of $\rm U=50$ ground users that are randomly distributed across the region of interest. For each constellation, we consider five gateways placed at real sites to reflect the actual ground footprint of the constellation. Satellite deployment and orbital motion are modelled using the \texttt{Skyfield} library with recent TLE data obtained from~\cite{celestrak}. The gateways are took into account to measure round trip time (RTT) and they operate with a minimum elevation angle of $10^\circ$ to reflect their high-gain, large-aperture antennas. \texttt{Skyfield} enables for accurate estimation of the satellite positions over time. The user-satellite links are modelled using the channel that described in the system model of Section~\ref{subsec:systemmode}. The outage-aware transmit powers are computed for an outage probability of $\varepsilon=0.05$. The simulation parameters used throughout the paper are summarized in Table~\ref{table:parameters}~\cite{choi2025intelligent,11419899}.

The min-power, balanced, and inter-handover interval (max-IHI) GHAAP hyperparameter settings are set, respectively, to $(\alpha_1,\alpha_2) = (0.1,0.0)$, $(0.2,0.1)$, and $(0.4,0.7)$ for OneWeb, and to $(0.1,0.0)$, $(0.5,0.1)$, and $(1.0,0.7)$ for Starlink. The settings were selected from the Pareto front obtained via grid search. Additionally, we developed two risk-aware baselines, RA-MRSRP and RA-MST. These baselines obtained by utilizing the proposed risk power consumption into two conventional association counterparts. In other words, both baselines share the same VaR-derived transmit power $p_{u,s}^{\rm min} (t,{\varepsilon})$ obtained from the power allocation subproblem.  
Each user is associated with the satellite has the strongest nominal received signal. Accordingly, the satellites are sorted in decreasing signal order. Then, the user is admitted to the first satellite, that satisfy the power budget and the free-beam constraints. Additionally, users are sorted in increasing order based on the power consumption, so that the users with lower transmit power are admitted first to maximizes the admitted user under the shared power and beam budget. Conversely, each user RA-MST is associated with the satellite of the longest remaining visibility duration subject to the VaR power budget and the beam cap. In RA-MST, a user keeps its current serving satellite as long as the link remains feasible. Therefore, re-associates happens only when the link is not feasible. Users are ordered in decreasing order of their remaining visibility, so the user is first associated with the link higher visibility duration. 
\begin{figure*}[t!]
\centering
\begin{subfigure}[b]{0.32\textwidth}
  \centering\includegraphics[width=\linewidth]{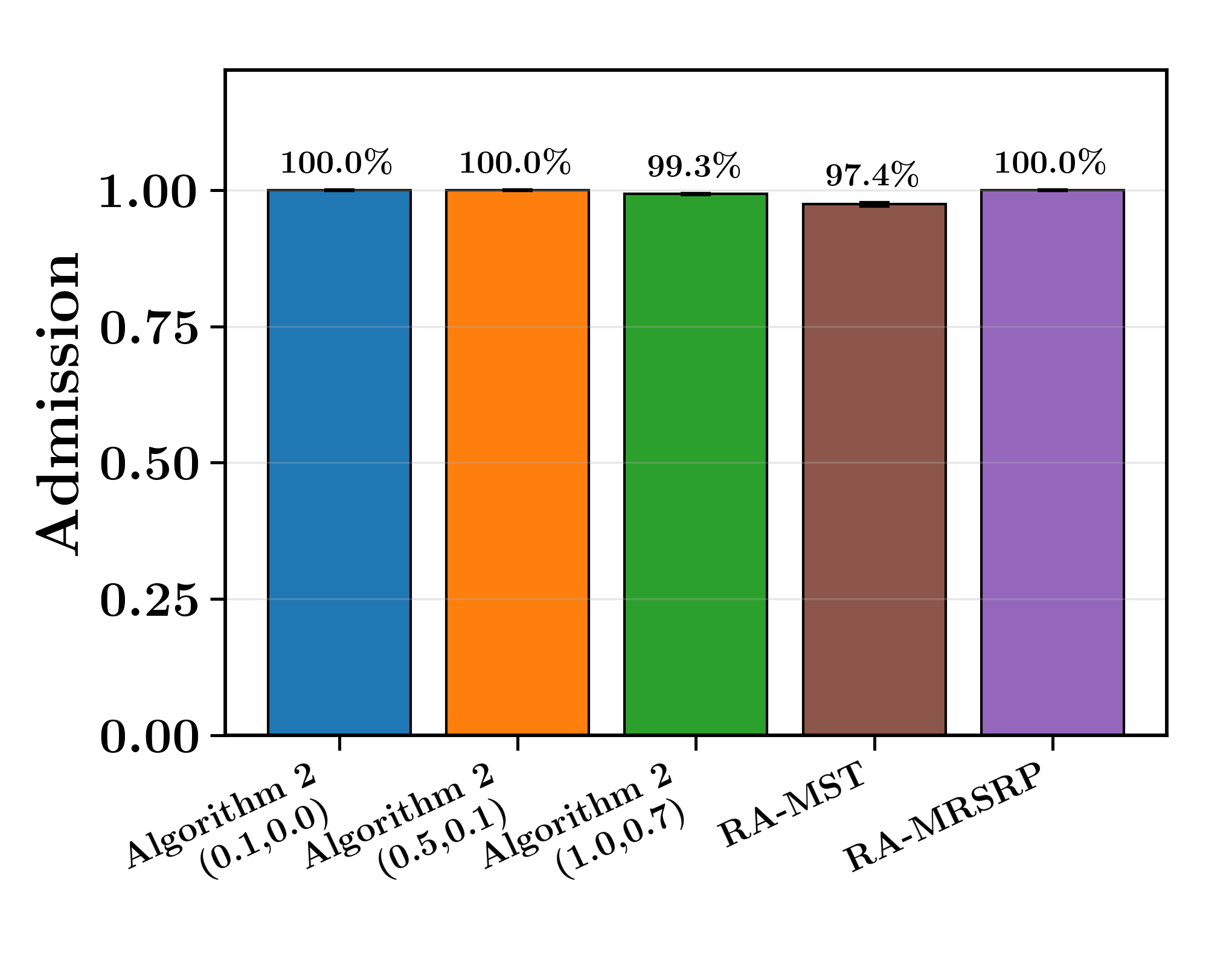}
  \caption{Starlink admission}\label{Fig:bar_star_admission}
\end{subfigure}\hfill
\begin{subfigure}[b]{0.32\textwidth}
  \centering\includegraphics[width=\linewidth]{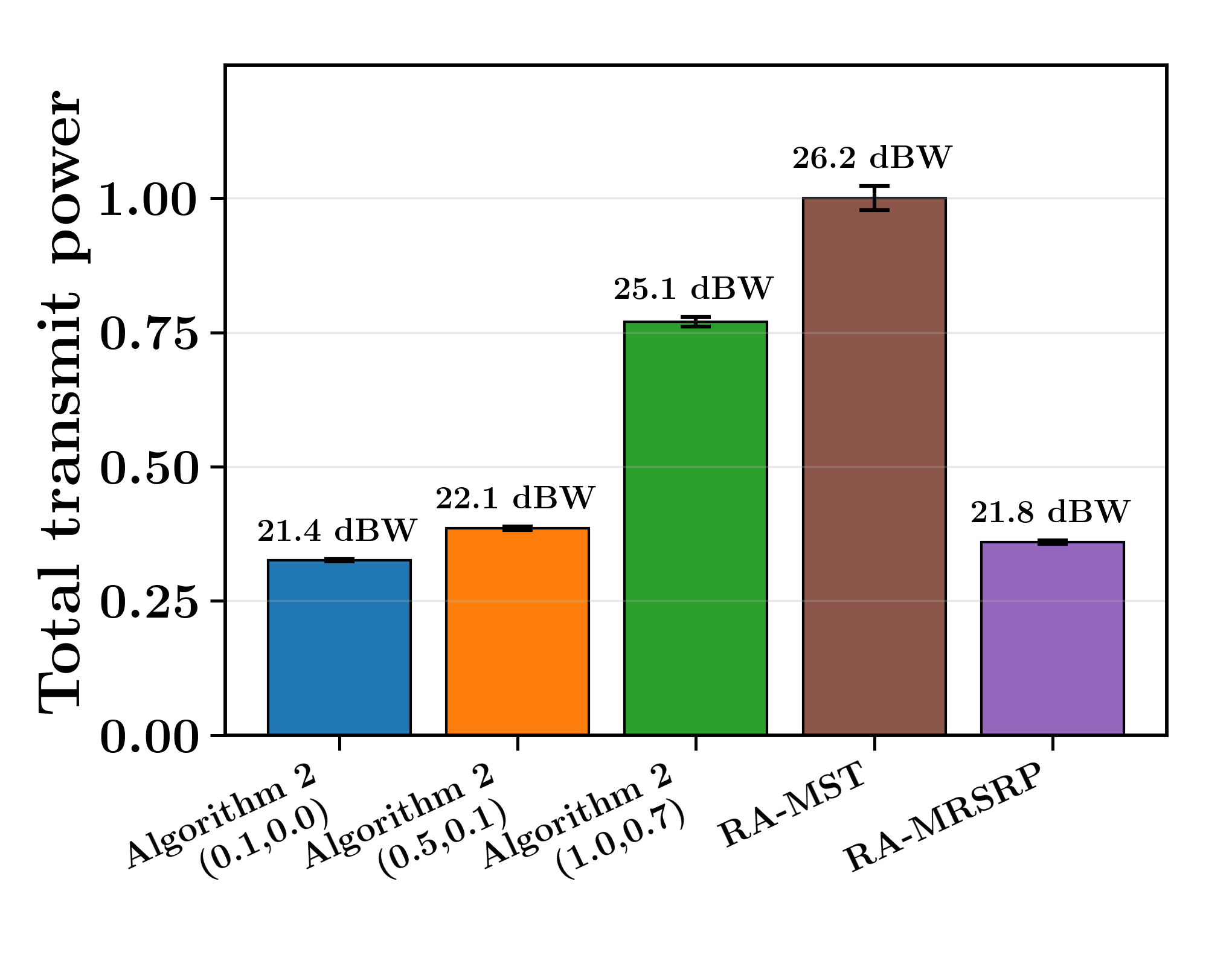}
  \caption{Starlink transmit power}\label{Fig:bar_star_power}
\end{subfigure}\hfill
\begin{subfigure}[b]{0.32\textwidth}
  \centering\includegraphics[width=\linewidth]{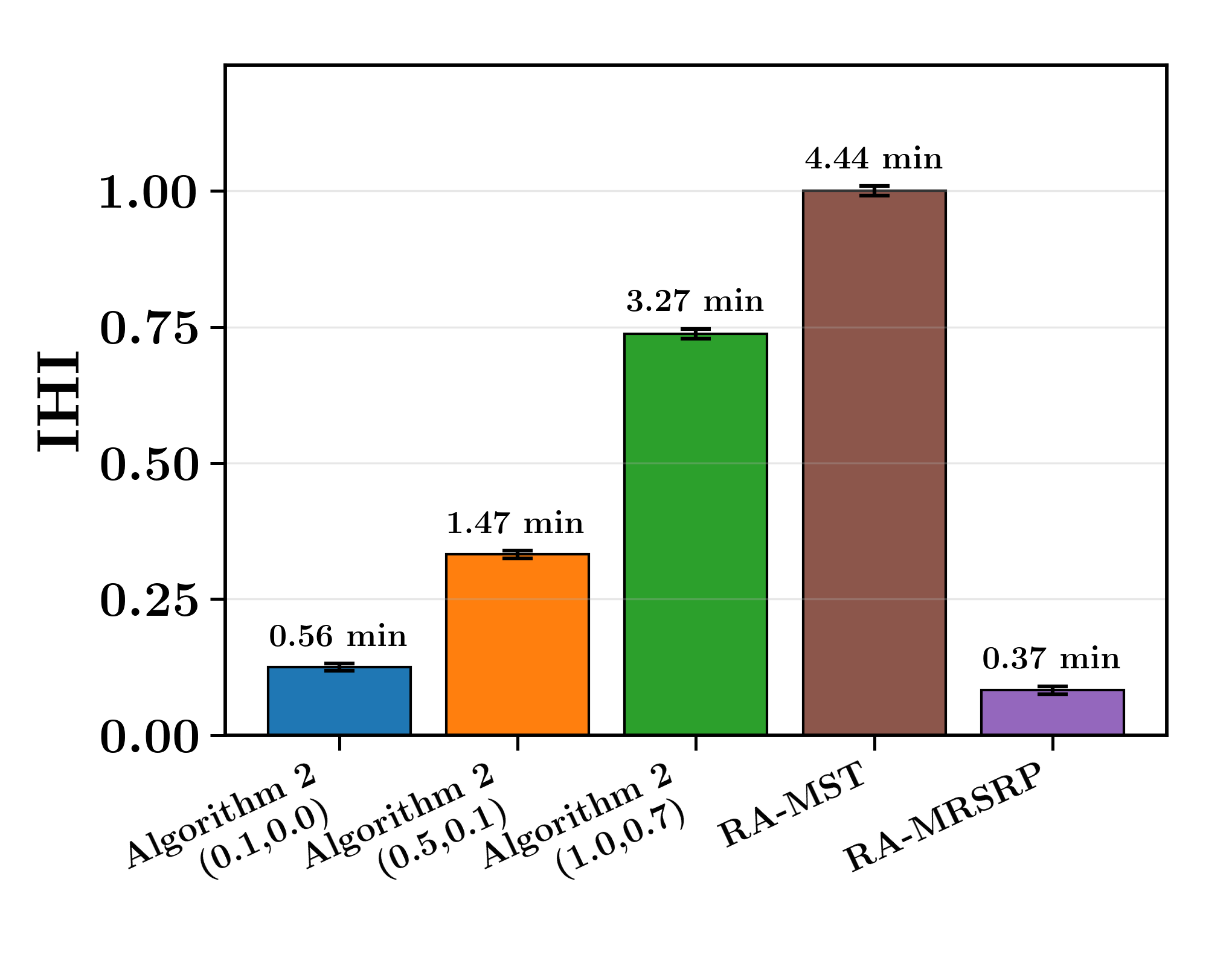}
  \caption{Starlink IHI}\label{Fig:bar_star_ihi}
\end{subfigure}

\vspace{-2pt}
\begin{subfigure}[b]{0.32\textwidth}
  \centering\includegraphics[width=\linewidth]{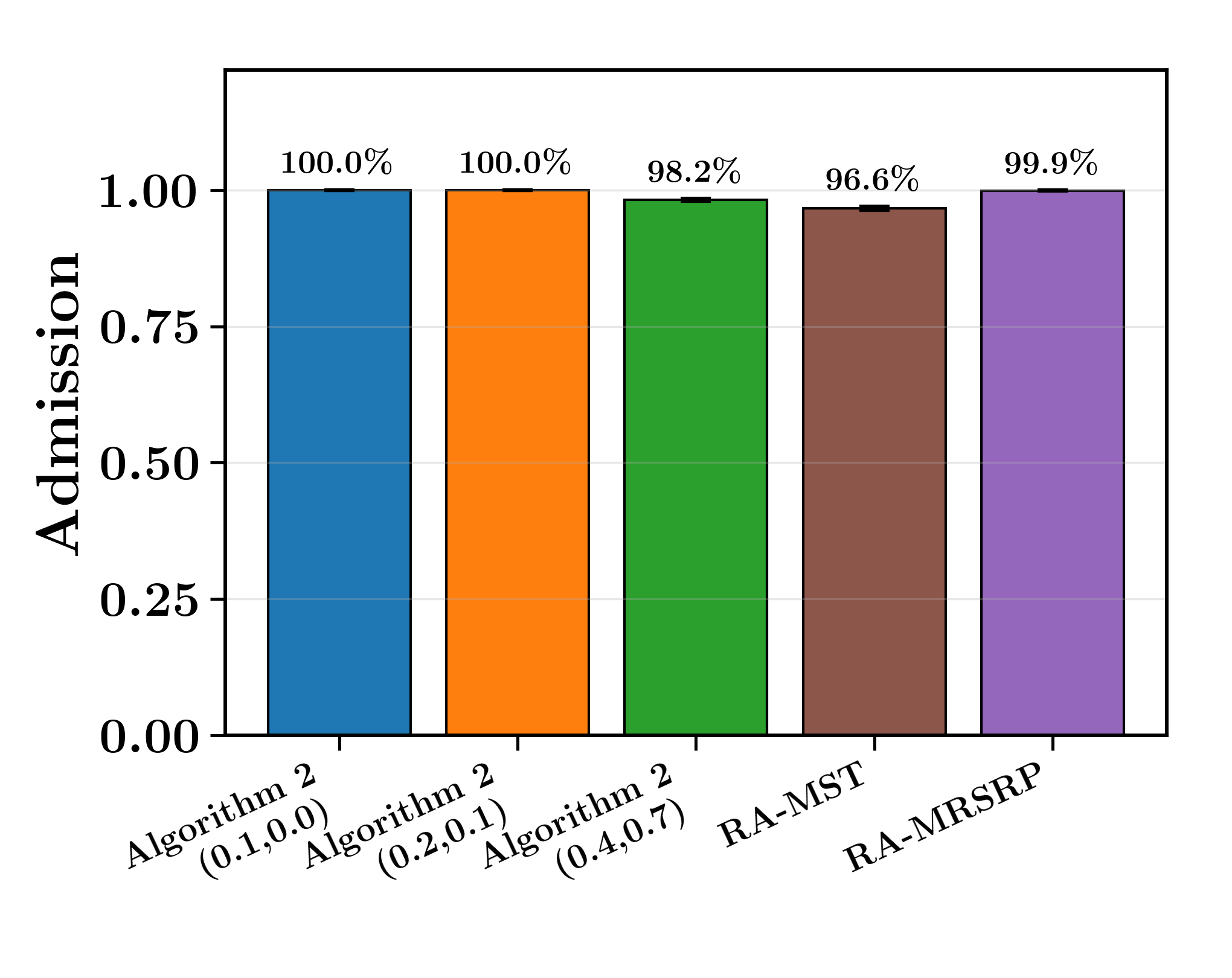}
  \caption{OneWeb admission}\label{Fig:bar_one_admission}
\end{subfigure}\hfill
\begin{subfigure}[b]{0.32\textwidth}
  \centering\includegraphics[width=\linewidth]{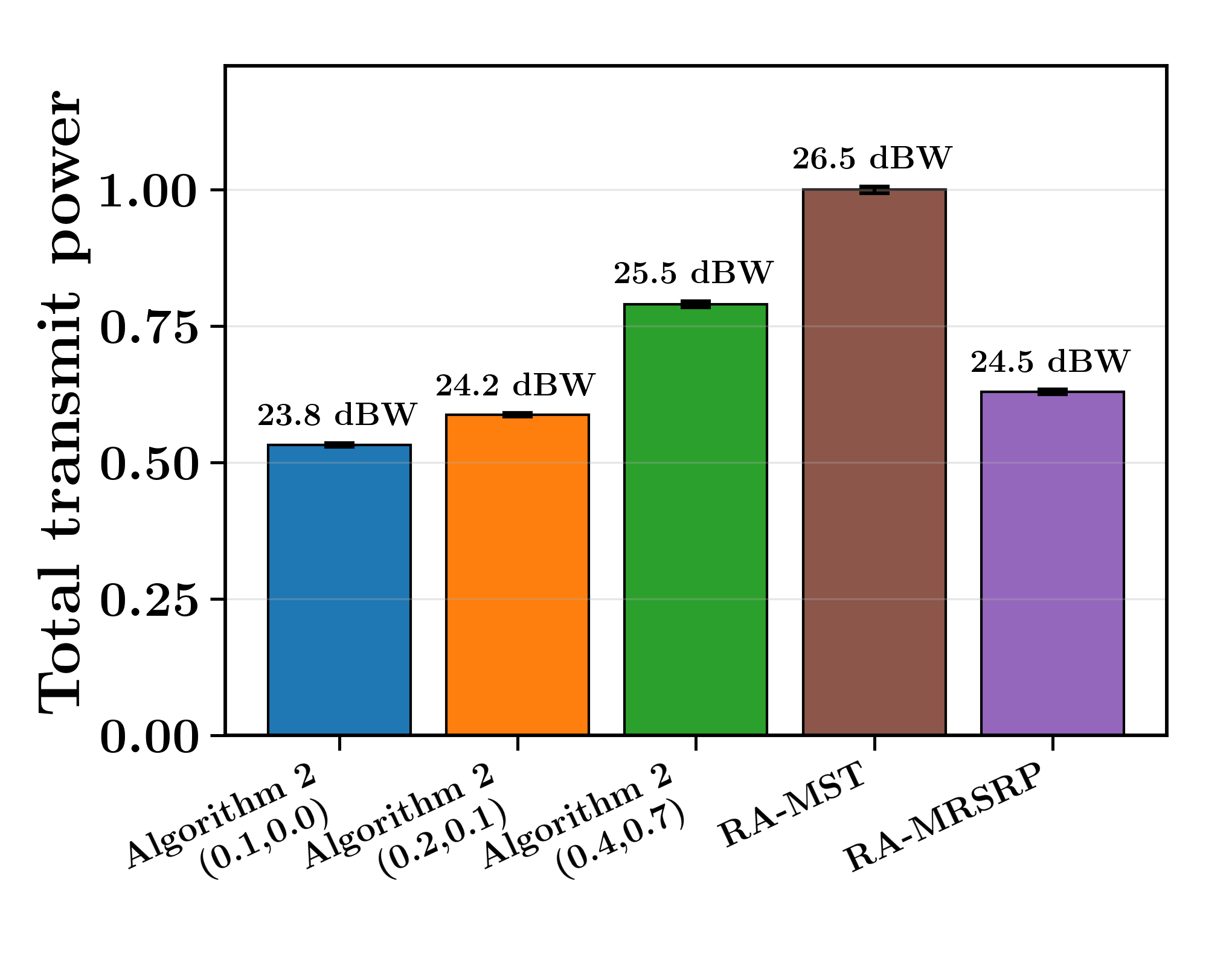}
  \caption{OneWeb transmit power}\label{Fig:bar_one_power}
\end{subfigure}\hfill
\begin{subfigure}[b]{0.32\textwidth}
  \centering\includegraphics[width=\linewidth]{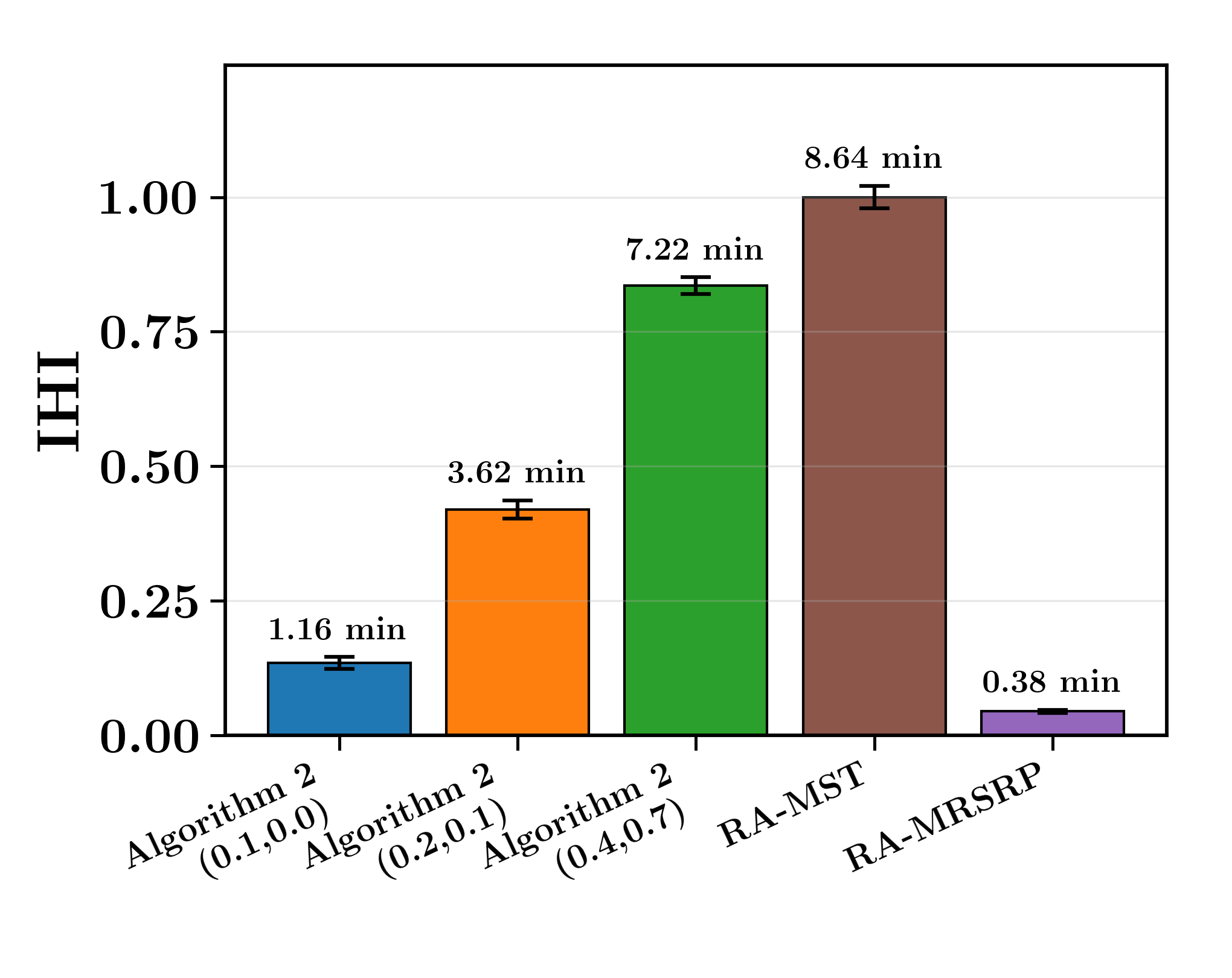}
  \caption{OneWeb IHI}\label{Fig:bar_one_ihi}
\end{subfigure}
\caption{Performance comparison for OneWeb and Starlink at spectral efficiency $2$~bps/Hz: admission, transmit power, and IHI.}
\label{fig:bar_compar}
\vspace{-0.2in}
\end{figure*}
\subsection{Performance Comparison of the Proposed Algorithms }
 Fig.~\ref{fig:optimality} compares the proposed Algorithms~1 and~2 for the three hyperparametes settings, i.e., transmit power, IHI, and admission. The comparison is performed while varying the spectral efficiency from $1$ to $5$~bps/Hz. In this context, the IHI is defined as the mean time between consecutive handovers of the serving satellite, averaged over all users and over the simulation time. The figure shows that Algorithm~2 achieves performance close to the ILP-based Algorithm~1. Moreover, the gap to the optimal solution narrows as the spectral efficiency requirement is relaxed, which confirms the effectiveness of Algorithm~2. In fact, the two solutions are practically similar in terms of transmit power and admission. At low spectral-efficiency requirements, the required transmit powers are small, and the per-satellite power-budget constraints are rarely active; hence, Algorithm~2 closely matches Algorithm~1. A small gap appears only at the highest spectral-efficiency requirement, where the larger required transmit powers make the per-satellite power budget the main limiting factor, causing Algorithm~2 to perform the knapsack-based repair stage more frequently. 
 The minimum-power setting attains the lowest transmit power, but at the expense of a shorter IHI, i.e., more frequent handovers. Conversely, the maximum-IHI setting achieves the best handover performance, although at the cost of a higher transmit power and a lower admission rate. For instance, at a spectral efficiency of $3$~bps/Hz, it yields an IHI of $9.78$~mins and a total transmit power of $27.89$~dBW, compared with $1.16$~mins and $26.89$~dBW for the minimum-power setting. Nevertheless, the power gap remains moderate, about $1.0$~dBW, while the admission rate drops by roughly $6.4\%$. The balancing point achieves an admission rate and a transmit power close to those of the minimum-power setting, while considerably increasing the IHI from $1.16$ to $4.02$~mins, i.e., reduced handover frequency.\vspace{-.1 in}

\begin{figure}
\centering
\begin{subfigure}[b]{0.8\columnwidth}
  \centering\includegraphics[width=\linewidth]{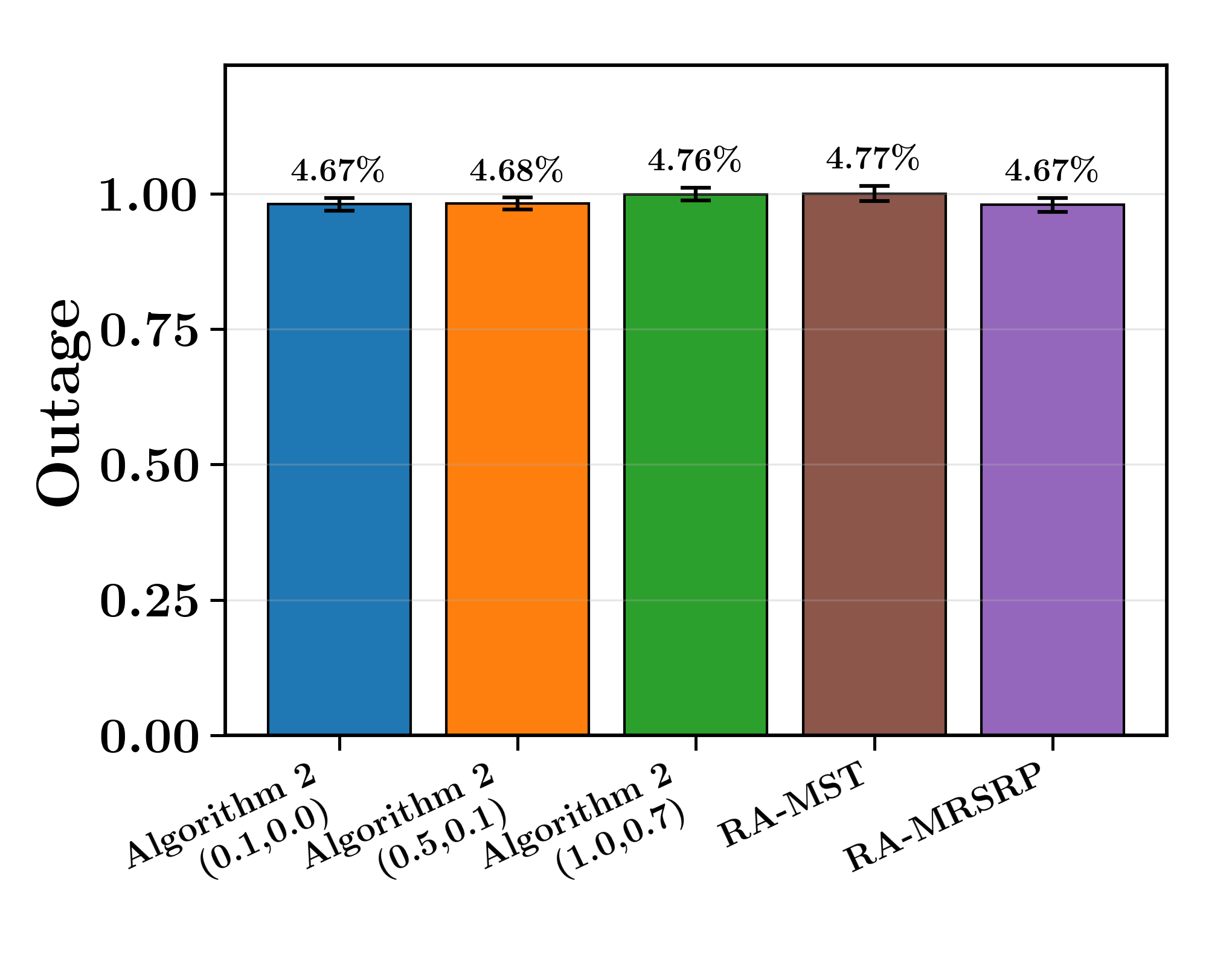}
  \caption{Starlink outage}\label{Fig:bar_star_out}
\end{subfigure}\\
\vspace{-0pt}
\begin{subfigure}[b]{0.8\columnwidth}
  \centering\includegraphics[width=\linewidth]{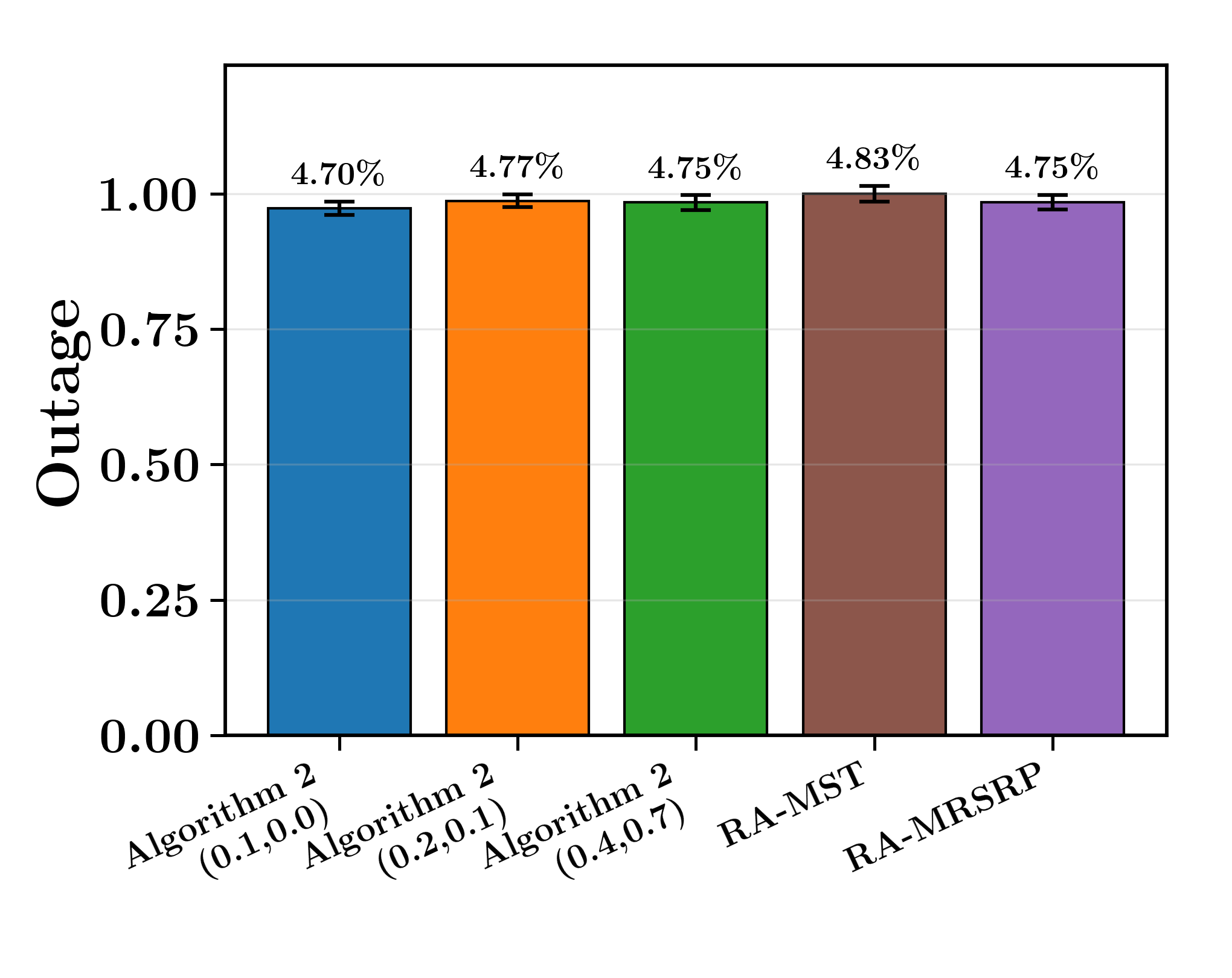}
  \caption{OneWeb outage}\label{Fig:bar_one_out}
\end{subfigure}
\caption{Outage of the proposed schemes for Starlink and OneWeb at spectral efficiency $2$~bps/Hz.}
\label{fig:bar_outage}
\vspace{-0.0in}
\end{figure}

\begin{figure*}[t!]
\centering
\captionsetup[subfigure]{font=footnotesize}
\begin{subfigure}[b]{0.32\textwidth}
  \centering\includegraphics[width=\linewidth]{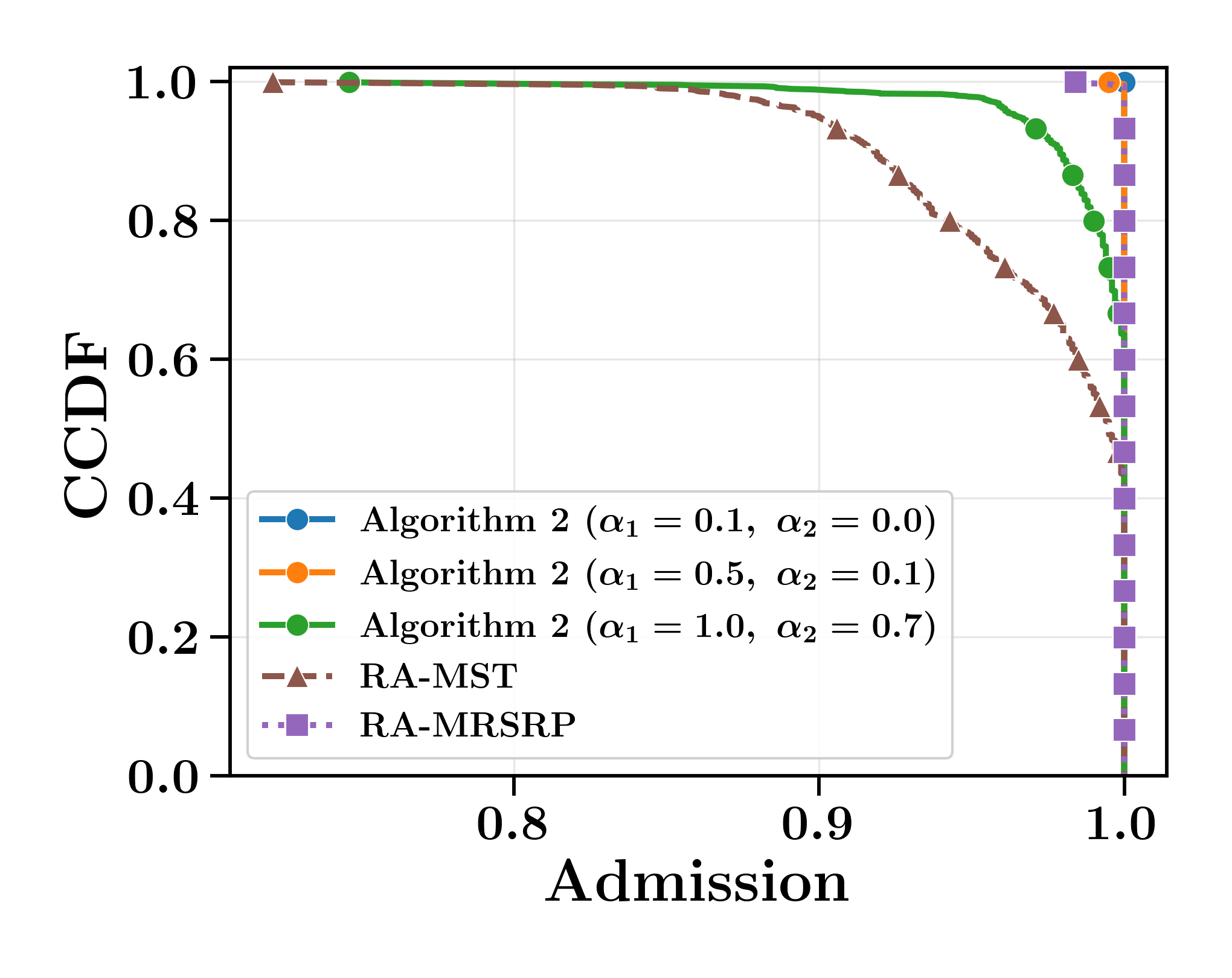}
  \caption{Starlink admission}\label{Fig:ccdf_star_admission}
\end{subfigure}\hfill
\begin{subfigure}[b]{0.32\textwidth}
  \centering\includegraphics[width=\linewidth]{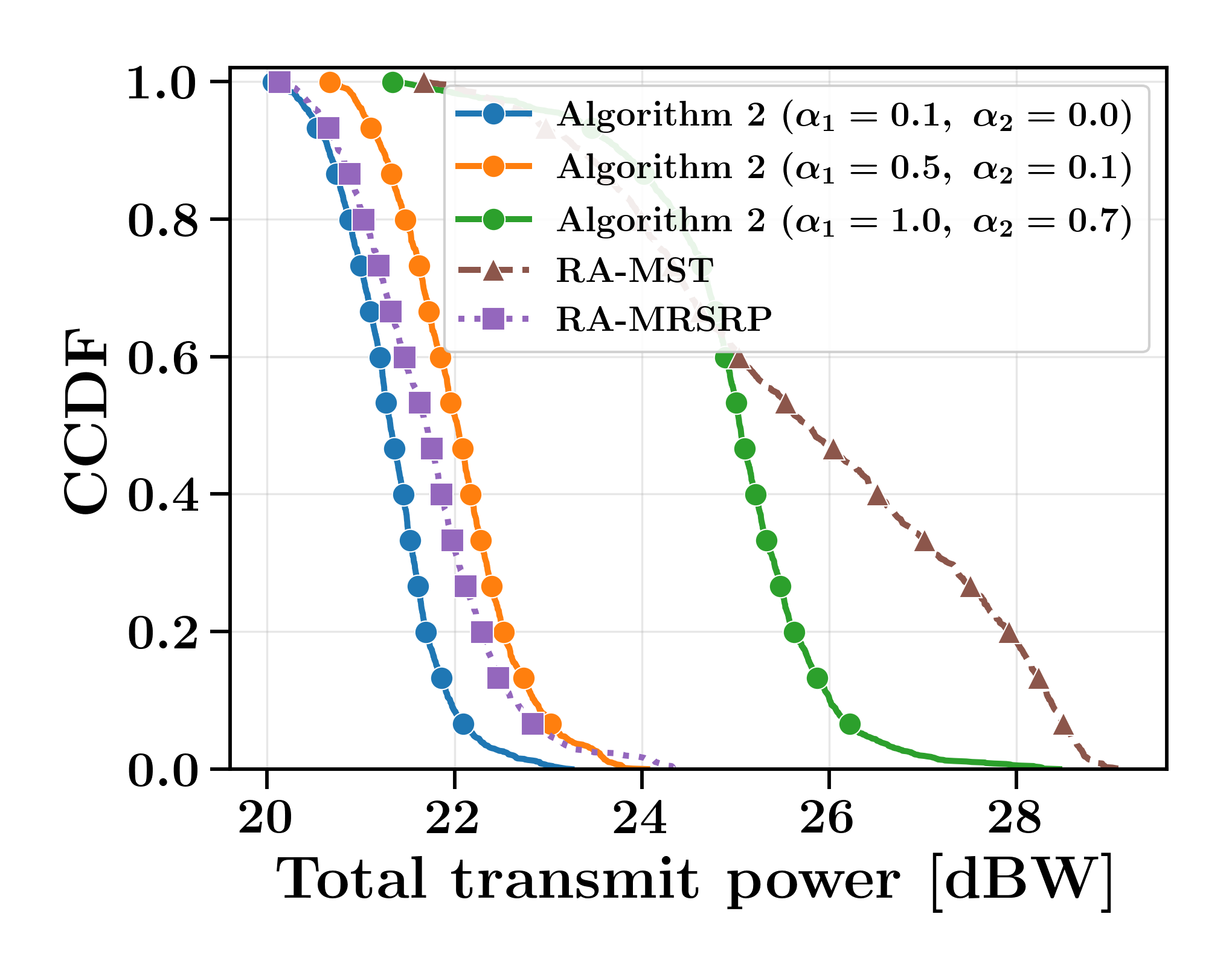}
  \caption{Starlink transmit power}\label{Fig:ccdf_star_power}
\end{subfigure}\hfill
\begin{subfigure}[b]{0.32\textwidth}
  \centering\includegraphics[width=\linewidth]{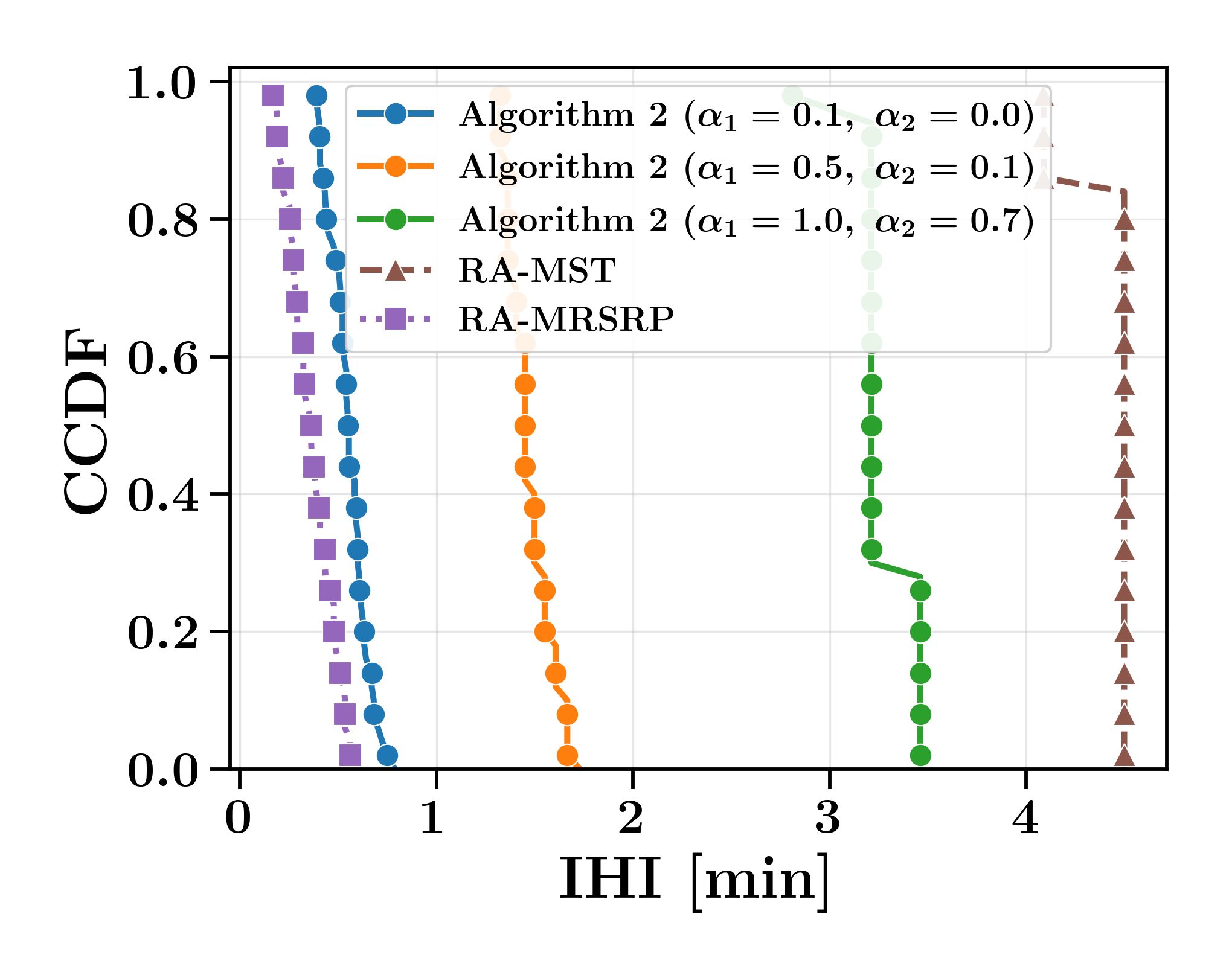}
  \caption{Starlink IHI}\label{Fig:ccdf_star_ihi}
\end{subfigure}

\vspace{-2pt}
\begin{subfigure}[b]{0.32\textwidth}
  \centering\includegraphics[width=\linewidth]{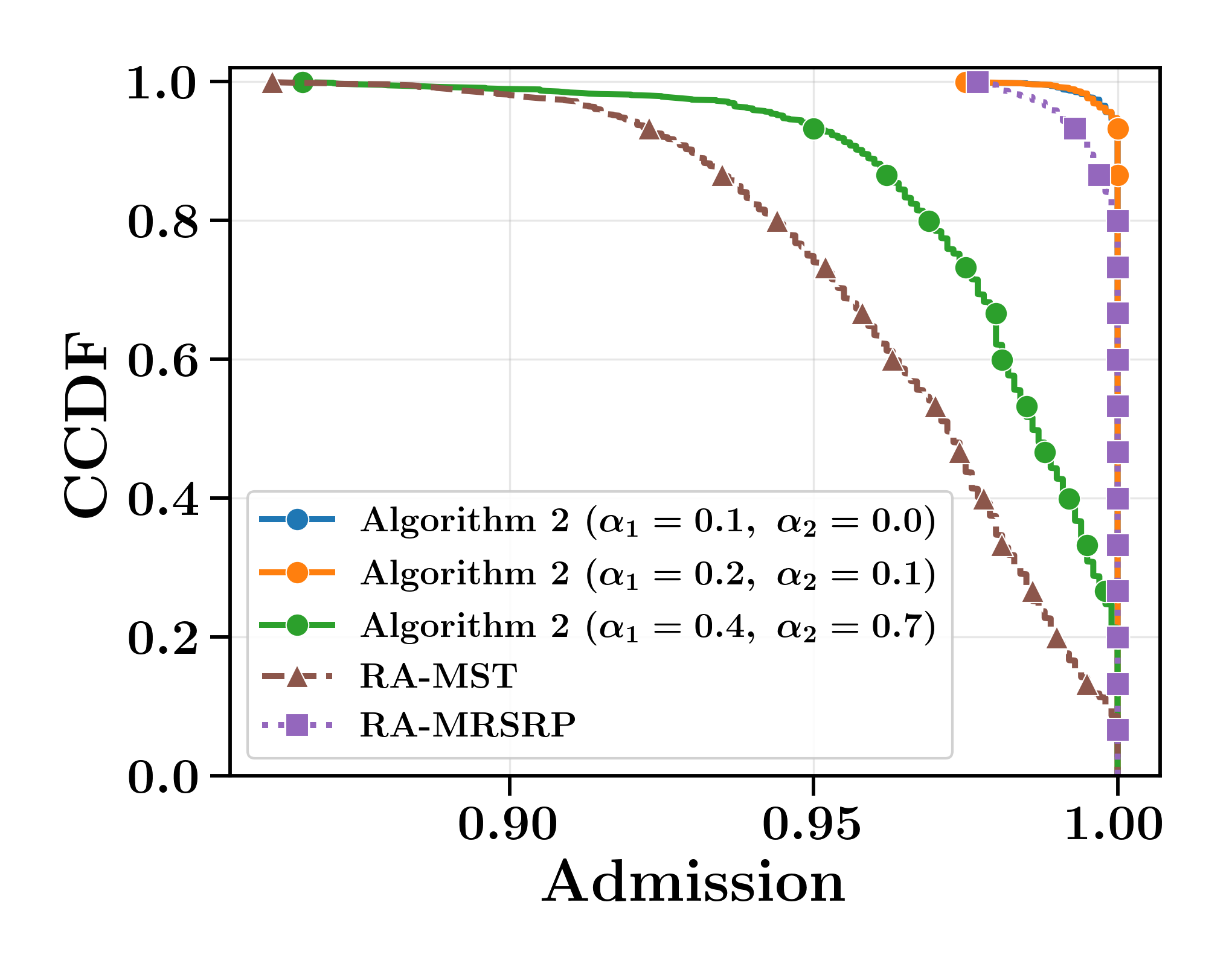}
  \caption{OneWeb admission}\label{Fig:ccdf_one_admission}
\end{subfigure}\hfill
\begin{subfigure}[b]{0.32\textwidth}
  \centering\includegraphics[width=\linewidth]{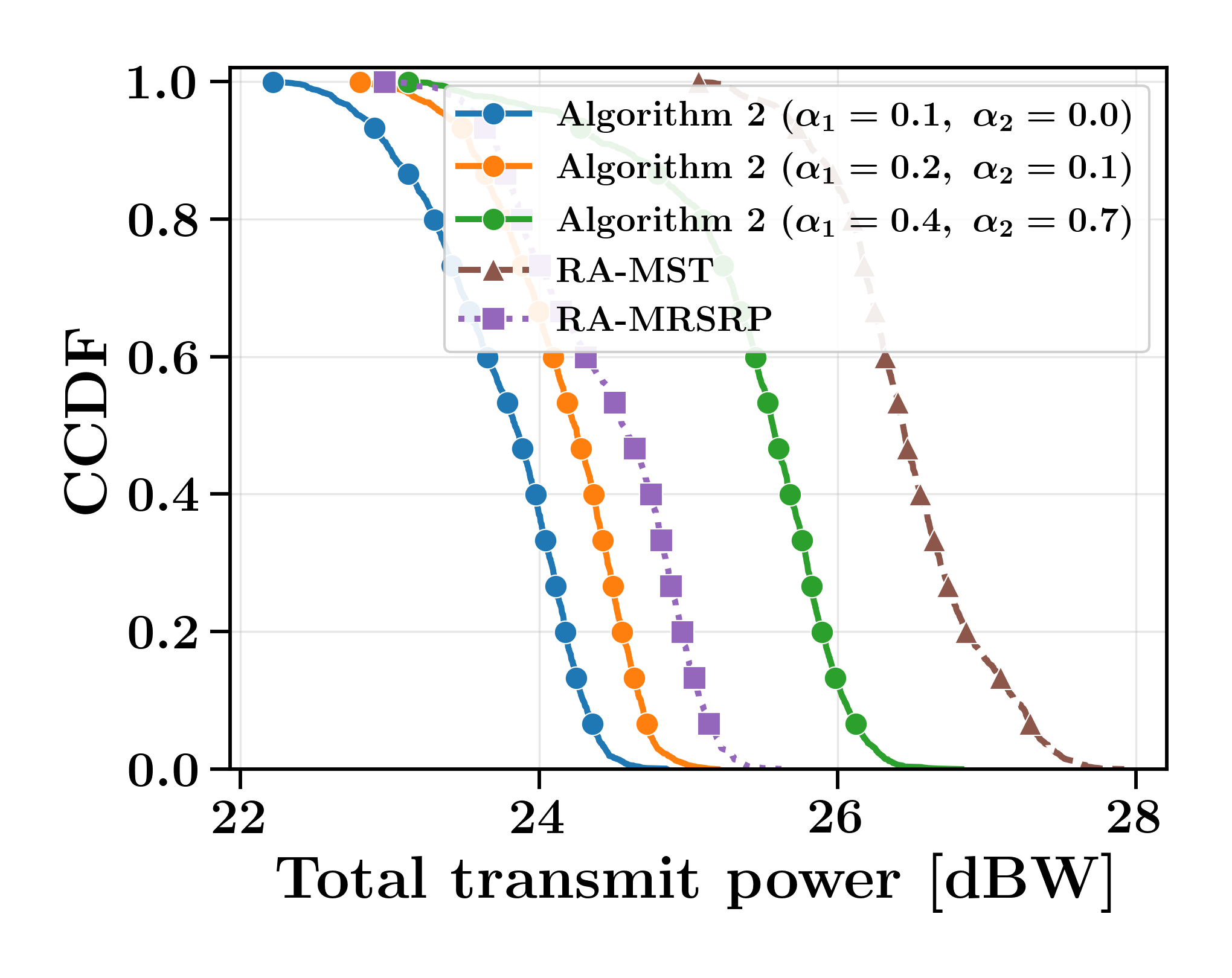}
  \caption{OneWeb transmit power}\label{Fig:ccdf_one_power}
\end{subfigure}\hfill
\begin{subfigure}[b]{0.32\textwidth}
  \centering\includegraphics[width=\linewidth]{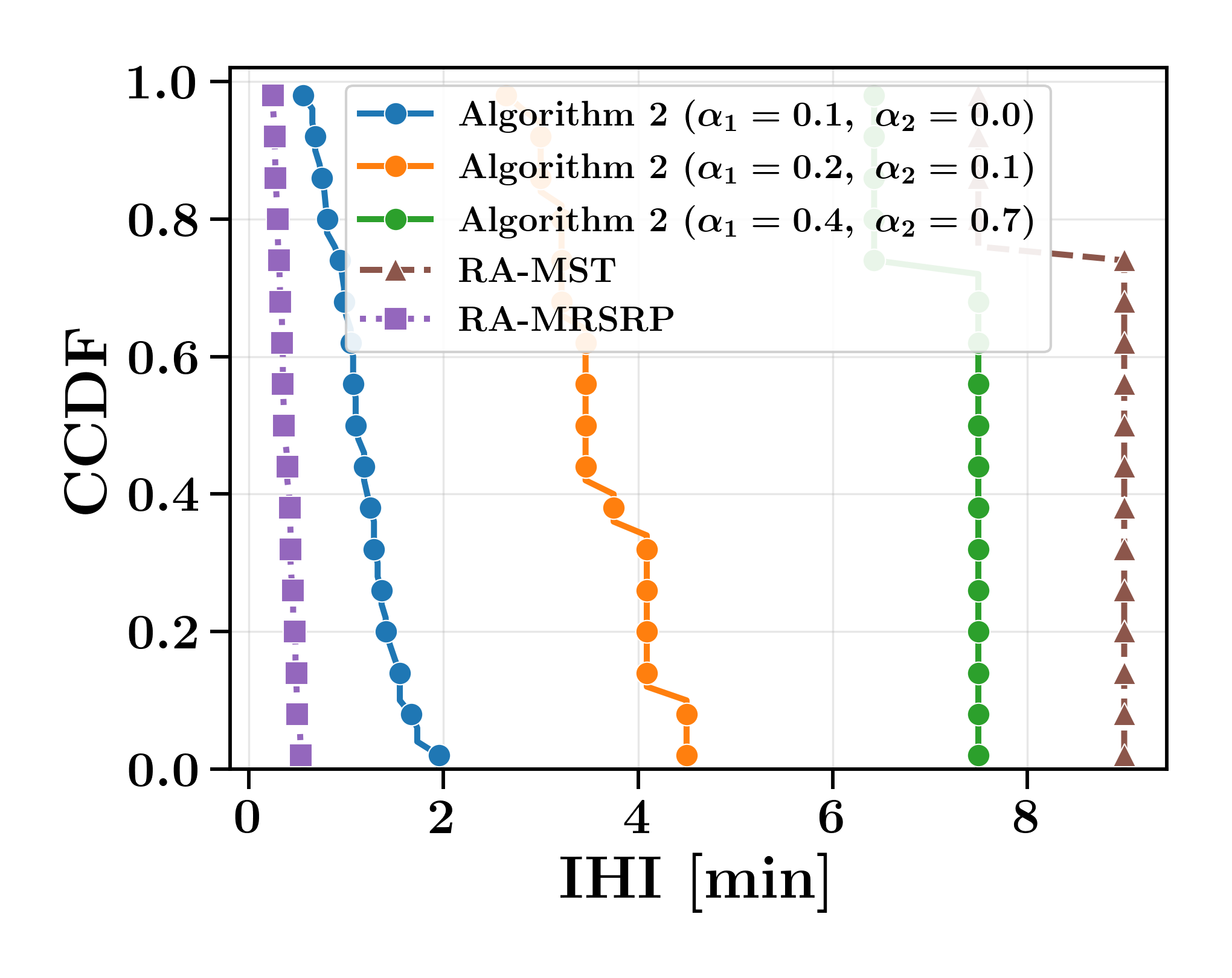}
  \caption{OneWeb IHI}\label{Fig:ccdf_one_ihi}
\end{subfigure}
\caption{Distribution comparison for OneWeb and Starlink at spectral efficiency $2$~bps/Hz: admission, transmit power, and IHI.}
\label{fig:ccdf_compar}
\vspace{-0.2in}
\end{figure*}

\subsection{Metrics Performance}
Fig.~\ref{fig:bar_compar} presents a performance comparison of the proposed scheme and the risk-aware baselines for both the Starlink and OneWeb constellations, in terms of transmit power, user admission, and IHI, at a spectral efficiency of $2$~bps/Hz. This operating point is carefully selected to give fair results for the two constellations without biasing the comparison toward either of them. For both constellations, all schemes exhibit the same trend for the three metrics. The two constellations differ mainly in scale. Specifically, OneWeb have roughly double the IHI of Starlink for all schemes. For example, the balanced setting grows from $1.47$ to $3.62$~mins and the maximum-IHI setting from $3.27$ to $7.22$~mins, because its higher orbit means longer satellite visibility. The only exception is RA-MRSRP, whose IHI is essentially unchanged ($0.37$ against $0.38$~mins). This is because of that it always re-associates to the nearest satellite, which means frequent handovers despite the orbital altitude. Similarly, Starlink consumes less transmit power than OneWeb, by up to about $2.7$~dBW. This gap is maximum for the schemes that connect to nearby satellites, such as RA-MRSRP and the minimum-power setting, and reduced to about $0.3$~dBW for RA-MST. This behaviour is expected, since RA-MST is less sensitive to the altitude difference between the two constellations.

As expected, RA-MRSRP achieves the lowest transmit power among the baselines ($21.8$~dBW on Starlink) because it always connects to the satellite providing the highest RSRP, i.e., satellite with lower slant range. However, this low power consumption comes at the expense of the highest handover rate, i.e., lowest IHI ($0.37$~mins). Oppositely, RA-MST exhibits the highest transmit power ($26.2$~dBW) while achieving the lowest handover rate, i.e., the highest IHI ($4.44$~mins). This is because that RA-MST gives higher priority for satellites with longer visibility duration. Additionally, it reduces the admission to $97.4\%$, as it tends to consume more transmit power of the satellite budget. In contrast, the proposed scheme provides a balanced trade-off between transmit power and handover frequency while maintaining an admission probability close to that of RA-MRSRP, i.e., about $99$--$100\%$. By tuning $(\alpha_1,\alpha_2)$ it spans the entire operating range: at its minimum-power setting it improves on RA-MRSRP by $0.4$~dBW while remaining QoS-guaranteed, and at its maximum-IHI setting it reaches an IHI ($3.27$~mins) close to that of RA-MST while consuming about $1.1$~dBW less power and admitting more users ($99.3\%$ against $97.4\%$). Fig.~\ref{fig:bar_outage} shows that the outage remains below the target risk level $\epsilon=5\%$ for every scheme and every operating point (between $4.67\%$ and $4.83\%$). This confirms that the proposed composite log-normal approximation guarantees the target reliability, independent of the association algorithm and the constellation.

\begin{figure}[t!]
\centering
\captionsetup[subfigure]{font=footnotesize}
\begin{subfigure}[b]{0.8\columnwidth}
  \centering\includegraphics[width=\linewidth]{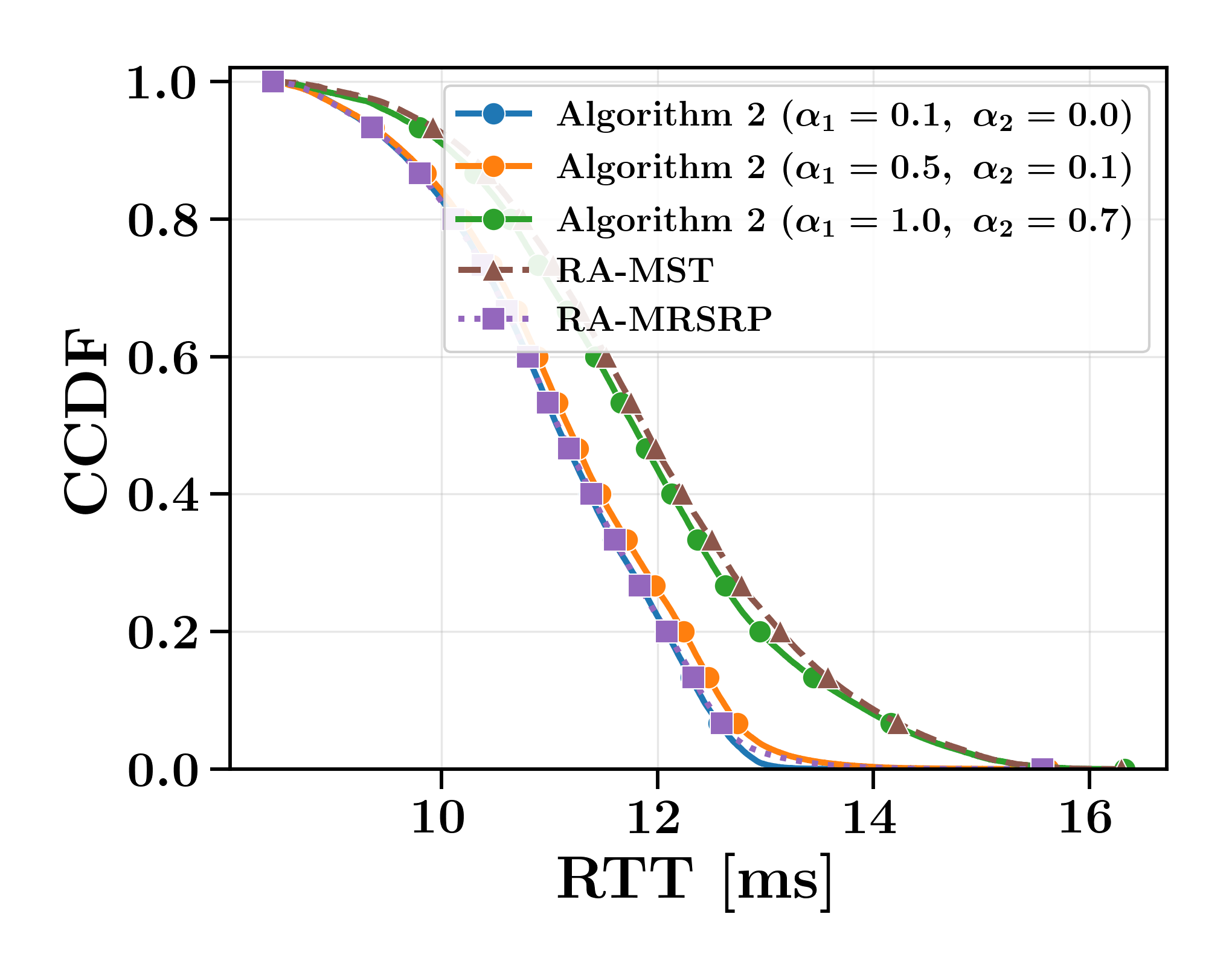}
  \caption{Starlink RTT}\label{Fig:ccdf_star_rtt}
\end{subfigure}\\
\vspace{-2pt}
\begin{subfigure}[b]{0.8\columnwidth}
  \centering\includegraphics[width=\linewidth]{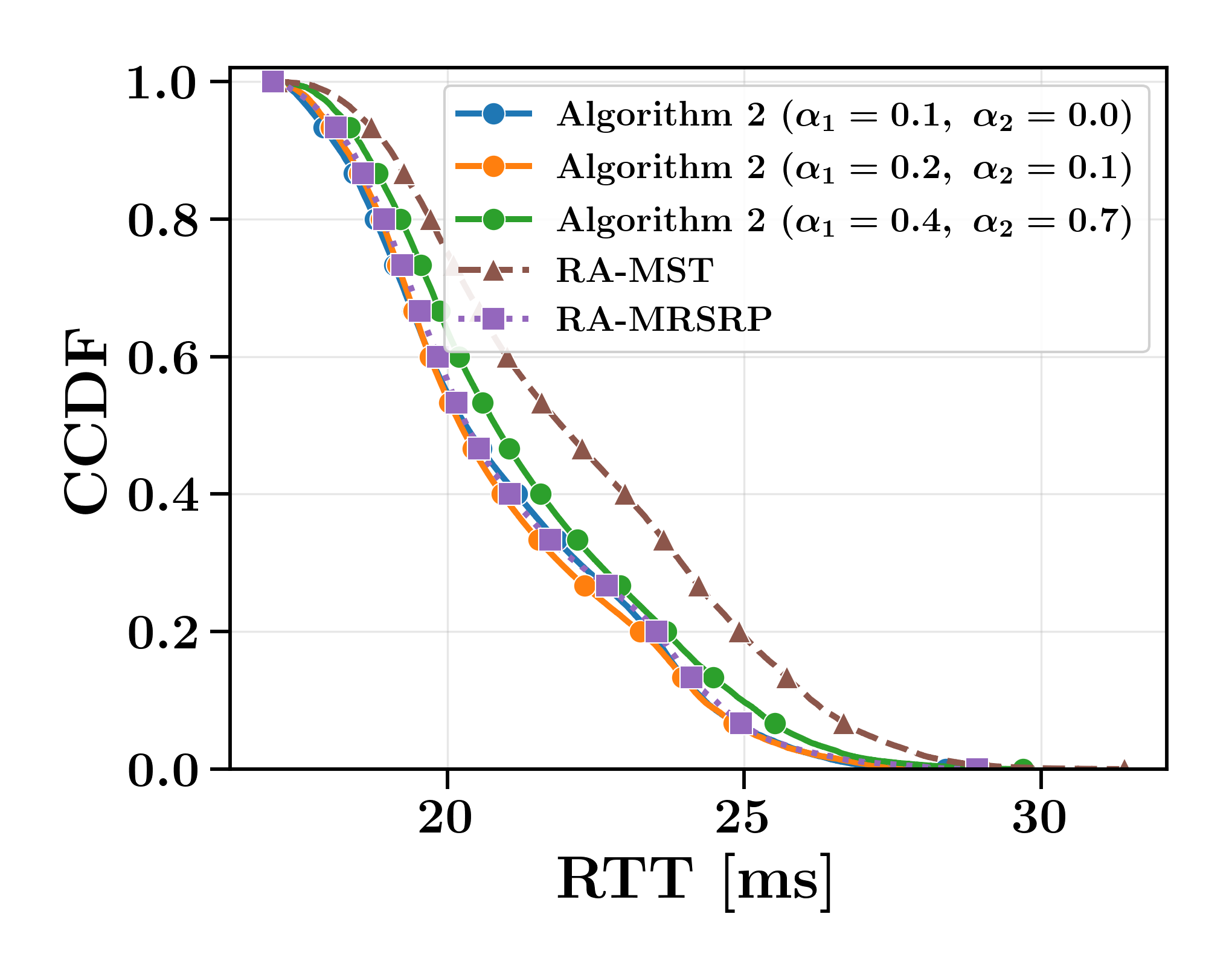}
  \caption{OneWeb RTT}\label{Fig:ccdf_one_rtt}
\end{subfigure}
\caption{Round-trip time (RTT) distribution for Starlink and OneWeb at spectral efficiency $2$~bps/Hz.}
\label{fig:ccdf_rtt}
\vspace{-0.2in}
\end{figure}

Fig.~\ref{fig:ccdf_compar} depicts the CCDFs of the users' admission, transmit power, and IHI for both Starlink and OneWeb constellations at $2$~bps/Hz. It shows that RA-MRSRP and the minimum-power setting of the proposed scheme are concentrated at low power. For Starlink, Fig.~\ref{Fig:ccdf_star_power} illustrates that the median of both solutions is about $21.5$~dBW and the $95^{th}$ percentile of only $22.2$--$23.0$~dBW. This means that they not only minimize the transmit power but also stable the power across links. In contrast, RA-MST and the maximum-IHI setting exhibit longer tails, hence, the $95^{th}$ percentile reaching $28.6$ and $26.4$~dBW respectively. This indicates that links varies widely with respect the path loss. The same behaviour is observed on OneWeb. The CCDFs of the IHI are clearly separated and follow the order that RA-MRSRP $<$ minimum-power $<$ balanced $<$ maximum-IHI $<$ RA-MST. Additionally, OneWeb distribution is wider, roughly twice the Starlink values, except for RA-MRSRP as it tends to concentrate at low power consumptions. Fig.~\ref{fig:ccdf_rtt} shows that the RTT CCDFs are dominated by the constellation geometry. In this regards, the RTT is calculated to nearest gateway. OneWeb RTTs are about twice the Starlink values, with a median of about $ 20$~ms against about $ 11$~ms. This due to the higher orbit and the longer feeder link to the nearest gateway. Similarly, RA-MST and the maximum-IHI setting showing a slightly larger tail, up to $14.5$~ms on Starlink and $27$~ms on OneWeb. Finally, RA-MST exhibits a longer tail, because it consumes more power, while the proposed maximum-IHI setting shows only a mild tail down to $0.75$. This confirms that the proposed scheme sustains a high and stable admission while offering a controllable power/handover trade-off.

\section{Conclusion}
\label{sec:conclusion}
This work proposes a risk-aware handover management framework for low Earth orbit (LEO) satellite networks. The problem is formulated as a mixed-integer nonlinear program that jointly minimizes transmit power consumption and handover rate while maximizing the number of served users under per-satellite power budgets and stochastic quality-of-service (QoS) chance constraints. To control the handovers, we consider a generalized handover-aware association preference (GHAAP) score that consider the the distance between the satellite and users, expected visibility duration, and the age of the current connection.
Because the problem is NP-hard, we decompose it into an inner power allocation and outer generalized assignment subproblems. The minimum transmit power of the inner problem is obtained in closed form by adopting value-at-risk (VaR) to reformulate the chance constraint. The outer mixed-integer subproblem is solved to optimality using off-the-shelf solver. Then, we propose a low-complexity algorithm combining Hungarian assignment with per-satellite knapsack algorithms. In addition, we introduced \texttt{NTN-Connect}, a modular, configuration-driven, and reproducible platform for large-scale non-terrestrial networks (NTNs). Extensive simulations demonstrated that the proposed algorithm achieves near-optimal performance with polynomial complexity. Numerical results using \texttt{NTN-Connect} shows a balanced trade-off among admission rate, transmit power, and handover rate compared with the risk-aware maximum stay (RA-MST) and risk-aware maximum received signal received power (RA-MRSRP) baselines. 
For future work, we will focus on adaptive hyperparameter tuning of the GHAAP score, where the preference parameters and the admission reward are adjusted according to the instantaneous network conditions, allowing the operating point to track the interplay between the power consumption, handovers, and admission of users.

\bibliographystyle{IEEEtran}
\appendices

\section{Proof of Proposition 1}
\label{app:mgf}

\begin{proof}
For any user--satellite link, the chance constraint \eqref{eq:P1-qos} is equivalent to

\begin{equation}
\Pr\!\left(
|h|^{2}
\ge
\frac{\gamma^{\rm th}(\beta+\sigma_n^2)}
{p}
\right)
\ge
1-\epsilon.
\label{eq:proof1}
\end{equation}
Let $
|h|^{2}=\bar g\,Z,$ where $\bar g$ denotes the deterministic channel gain and $Z$ the normalized composite fading coefficient. Then, the chance constraint in~\eqref{eq:proof1} becomes
\begin{equation}
\Pr(Z\ge\gamma)\ge1-\epsilon,
\label{eq:proof2}
\end{equation}
where
\begin{equation}
\gamma
\triangleq
\frac{\gamma^{\rm th}(\beta+\sigma_n^2)}
{p\bar g}.
\label{eq:gamma_norm}
\end{equation}
Because $Z$ is a continuous random variable, $\Pr(Z\ge\gamma)=1-F_Z(\gamma)$, where $F_Z(\cdot)$ denotes the cumulative distribution function of $Z$. Then,
\begin{equation}
\Pr(Z\ge\gamma)\ge1-\epsilon
\iff
\gamma\le F_Z^{-1}(\epsilon),
\label{eq:proof3}
\end{equation}
where $F_Z^{-1}(\epsilon)$ denotes the lower $\epsilon$-quantile of the composite fading random variable. Let
$
\Psi_Z(\lambda)
\triangleq
\mathbb{E}\!\left[e^{-\lambda Z}\right]
$
denote the Laplace transform of the composite fading random variable $Z$. The composite lognormal--Rice fading distribution does not admit a tractable closed-form expression for its cumulative distribution function. Following the equivalent-lognormal approximation of the lognormal--Rice distribution in~\cite{4275022}, the composite fading variable $Z$ is approximated by an equivalent lognormal random variable $Y$. Let
$
Y
\triangleq
10^{X_Y/10},
\text{ where }
X_Y\sim\mathcal N(\mu_Y,\sigma_Y^2),
$
whose Laplace-form MGF is given by
\begin{equation}
\Psi_Y(\lambda;\mu_Y,\sigma_Y)
=
\sum_{n=1}^{N}
\frac{w_n}{\sqrt{\pi}}
\exp\!\left(
-\lambda
e^{(\sqrt2\sigma_Ya_n+\mu_Y)/\xi}
\right).
\label{eq:mgfY}
\end{equation}
where $\xi=\frac{10}{\ln10}, $ and $\{a_n,w_n\}$ denote the Gauss--Hermite abscissas and weights~\cite{4275022}. The logarithmic parameters $(\mu_Y,\sigma_Y)$ are obtained by numerically solving the two nonlinear MGF-matching equations
\begin{equation}
\Psi_Y(\lambda_i;\mu_Y,\sigma_Y)
=
\Psi_Z(\lambda_i),
\qquad
i=1,2,
\label{eq:mgfmatch}
\end{equation}
where $\lambda_1$ and $\lambda_2$ are prescribed matching points selected to improve the approximation accuracy over the region of interest of the distribution. Solving \eqref{eq:mgfmatch} determines the unknown parameters $(\mu_Y,\sigma_Y)$. In this work, the matching points are selected for the lower-tail
region associated with the outage constraint and their accuracy is
validated numerically against the exact lognormal--Rice CDF over the
considered channel-parameter range. 
Since $X_Y\sim\mathcal N(\mu_Y,\sigma_Y^2)$, the equivalent random variable $Y$ is lognormally distributed, whose cumulative distribution function is
\begin{equation}
F_Y(y)
=
\Phi\!\left(
\frac{10\log_{10}(y)-\mu_Y}
{\sigma_Y}
\right),
\label{eq:FY}
\end{equation}
where $\Phi(\cdot)$ denotes the cumulative distribution function of the standard normal distribution. Inverting \eqref{eq:FY} gives the lower $\epsilon$-quantile of the equivalent lognormal distribution,
\begin{equation}
F_Y^{-1}(\epsilon)
=
10^{(\mu_Y+\sigma_Y\Phi^{-1}(\epsilon))/10},
\label{eq:quantile2}
\end{equation}
where $\Phi^{-1}(\cdot)$ denotes the inverse cumulative distribution function of the standard normal distribution. Becuase the matched parameters $(\mu_Y,\sigma_Y)$ are functions of $K$ and the total shadowing standard deviation $\sigma_{\rm tot}$, define the equivalent-lognormal reserve margin as
$
D(\sigma_{\rm tot},K,\epsilon)
=
-
\left(
\mu_Y+\sigma_Y\Phi^{-1}(\epsilon)
\right).
$
As $
F_Z^{-1}(\epsilon)
\approx
F_Y^{-1}(\epsilon)
$, the lower $\epsilon$-quantile of the composite fading is approximated by
\begin{equation}
F_Z^{-1}(\epsilon)
\approx
10^{-D(\sigma_{\rm tot},K,\epsilon)/10}.
\label{eq:quantile3}
\end{equation}
Applying the approximation in \eqref{eq:quantile3} to \eqref{eq:proof3} yields
$
\gamma
\le
10^{-D(\sigma_{\rm tot},K,\epsilon)/10}
\label{eq:proof4}
$. Using the definition of $\gamma$ in \eqref{eq:gamma_norm} gives
\begin{equation}
p
\ge
\frac{
\gamma^{\rm th}(\beta+\sigma_n^2)
}
{
\bar g
10^{-D(\sigma_{\rm tot},K,\epsilon)/10}
}.
\label{eq:proof5}
\end{equation}
Because the objective is to minimizes the transmit power subject to the chance constraint, the lower bound in \eqref{eq:proof5} is attained at the optimum. Therefore,

\begin{equation}
p^{\min}(\epsilon)
=
\frac{
\gamma^{\rm th}(\beta+\sigma_n^2)
}
{
\bar g
}
10^{D(\sigma_{\rm tot},K,\epsilon)/10},
\label{eq:proof6}
\end{equation}
which completes the proof.
\end{proof}

\bibliography{refs}

\end{document}